\documentclass[%
 reprint,
 superscriptaddress,
preprintnumbers,
nofootinbib,
 amsmath,amssymb,
 aps,
 prd,
]{revtex4-1}

\usepackage{lineno}
\usepackage{tikz}
\usepackage{graphicx}% Include figure files
\usepackage{dcolumn}% Align table columns on decimal point
\usepackage{bm}% bold math
\usepackage{hyperref}% add hypertext capabilities
\usepackage{color}
\usepackage{array} %for space in a table
\usepackage{float} %for placing figures

\definecolor{airforceblue}{rgb}{0.36, 0.54, 0.66}
\definecolor{darkmidnightblue}{rgb}{0.0, 0.2, 0.4}

\definecolor{royalazure}{rgb}{0.0, 0.22, 0.66}
\definecolor{plasmablue}{rgb}{0.050383, 0.029803, 0.527975}
\usepackage{multirow, array}
\hypersetup{
    colorlinks=true,
    linkcolor=plasmablue,
    urlcolor=plasmablue,
    citecolor=plasmablue,
    linktoc=page,
}
\usepackage{aas_macros}

\renewcommand{\vec}{\bm}

\def\fun#1#2{\lower3.6pt\vbox{\baselineskip0pt\lineskip.9pt
        \ialign{$\mathsurround=0pt#1\hfill##\hfil$\crcr#2\crcr\sim\crcr}}}
\newcommand{\beq}{\begin{equation}}
\newcommand{\eeq}{\end{equation}}
\newcommand{\beqa}{\begin{eqnarray}}
\newcommand{\eeqa}{\end{eqnarray}}
\newcommand{\be}{\begin{equation}}
\newcommand{\ee}{\end{equation}}
\newcommand{\bea}{\begin{eqnarray}}
\newcommand{\eea}{\end{eqnarray}}
\newcommand{\nn}{\nonumber}

\newcommand{\dif}{\mathrm{d}}

\newcommand{\asz}{\ensuremath{{\zeta_0}}}
\newcommand{\bsz}{\ensuremath{{\zeta_\mathrm{M}}}}
\newcommand{\csz}{\ensuremath{{\zeta_\mathrm{z}}}}

\newcommand{\sigmalnzeta}{\ensuremath{{\sigma_{\ln\zeta}}}}
\newcommand{\alambda}{\ensuremath{{\lambda_0}}}
\newcommand{\blambda}{\ensuremath{{\lambda_\mathrm{M}}}}
\newcommand{\clambda}{\ensuremath{{\lambda_\mathrm{z}}}}

\newcommand{\sigmalnlambda}{\ensuremath{{\sigma_{\ln\lambda}}}}
\newcommand{\aWL}{\ensuremath{{\ln M_{\mathrm{WL}_0}}}}
\newcommand{\bWL}{\ensuremath{{M_{\mathrm{WL}_\mathrm{M}}}}}
\newcommand{\asigmaWL}{\ensuremath{{\ln\sigma^2_{\ln\mathrm{WL}_0}}}}
\newcommand{\bsigmaWL}{\ensuremath{{\sigma^2_{\ln\mathrm{WL}_\mathrm{M}}}}}

\newcommand{\Mwl}{\ensuremath{M_\mathrm{WL}}}
\newcommand{\Mhalo}{\ensuremath{M_\mathrm{halo}}} 
\newcommand{\Msun}{\ensuremath{\mathrm{M}_\odot}}

\newcommand{\Obhh}{\ensuremath{\Omega_\mathrm{b}h^2}}
\newcommand{\Om}{\ensuremath{\Omega_\mathrm{m}}}

\newcommand{\sig}{\ensuremath{\sigma_8}}
\newcommand{\Sopt}{\ensuremath{S_8^\mathrm{opt}}}
\newcommand{\LCDM}{\ensuremath{\Lambda\mathrm{CDM}}}

\definecolor{pygreen}{RGB}{44, 160, 44}
\usepackage[normalem]{ulem}

\makeatletter
\def\l@subsection#1#2{}
\def\l@subsubsection#1#2{}
\def\Dated@name{}
\makeatother

\begin{document}
\preprint{DES-2024-0859, FERMILAB-PUB-24-0859-PPD\\}
%\preprint{DES-2024-0859}
%\preprint{FERMILAB-PUB-24-0859-PPD}
\preprint{TUM-HEP-1541/24}

\title{Interacting Dark Sector (ETHOS $n=0$): Cosmological Constraints from SPT Cluster Abundance with DES and HST Weak Lensing Data}
\author{A.~Mazoun}
\email{asmaa.mazoun@tum.de}
\affiliation{Physik Department T31, Technische Universit\"at M\"unchen,
James-Franck-Stra\ss e 1, D-85748 Garching, Germany}
\affiliation{University Observatory, Faculty of Physics, Ludwig-Maximilians-Universit\"at, 
Scheinerstr. 1, D-81679 München, Germany}
\affiliation{Excellence Cluster ORIGINS, Boltzmannstr. 2, D-85748 Garching, Germany}
\author{S.~Bocquet}
%\email{}
\affiliation{University Observatory, Faculty of Physics, Ludwig-Maximilians-Universit\"at, Scheinerstr. 1, D-81679 München, Germany}

\author{J.~J.~Mohr}
%\email{}
\affiliation{University Observatory, Faculty of Physics, Ludwig-Maximilians-Universit\"at, Scheinerstr. 1, D-81679 München, Germany}
\affiliation{Max Planck Institute for Extraterrestrial Physics, Gießenbachstr.~1, D-85748 Garching, Germany}

\author{M.~Garny}
\affiliation{Physik Department T31, Technische Universit\"at M\"unchen,
James-Franck-Stra\ss e 1, D-85748 Garching, Germany}

\author{H.~Rubira}
\affiliation{Physik Department T31, Technische Universit\"at M\"unchen,
James-Franck-Stra\ss e 1, D-85748 Garching, Germany}
\affiliation{University Observatory, Faculty of Physics, Ludwig-Maximilians-Universit\"at, Scheinerstr. 1, D-81679 München, Germany}
\affiliation{Kavli Institute for Cosmology Cambridge, Madingley Road, Cambridge CB3 0HA, UK}
\affiliation{Department of Applied Mathematics and Theoretical Physics, University of Cambridge, Cambridge CB3 0WA, UK}
%\affiliation{Excellence Cluster ORIGINS, Boltzmannstr. 2, D-85748 Garching, Germany}

\author{M.~Klein}%\,\orcidlink{0000-0002-8248-4488}}
\affiliation{University Observatory, Faculty of Physics, Ludwig-Maximilians-Universit\"at, Scheinerstr. 1, D-81679 München, Germany}
\author{L.~E.~Bleem}%\,\orcidlink{0000-0001-7665-5079}}
\affiliation{High-Energy Physics Division, Argonne National Laboratory, 9700 South Cass Avenue, Lemont, IL 60439, USA}
\affiliation{Kavli Institute for Cosmological Physics, University of Chicago, 5640 South Ellis Avenue, Chicago, IL 60637, USA}

\author{S.~Grandis}%\,\orcidlink{0000-0002-4577-8217}}
\affiliation{Universit\"at Innsbruck, Institut f\"ur Astro- und Teilchenphysik, Technikerstr. 25/8, 6020 Innsbruck, Austria}

\author{T.~Schrabback}%\,\orcidlink{0000-0002-6987-7834}}
\affiliation{Universit\"at Innsbruck, Institut f\"ur Astro- und Teilchenphysik, Technikerstr. 25/8, 6020 Innsbruck, Austria}
\affiliation{Argelander-Institut f\"ur Astronomie, Auf dem H\"ugel 71, 53121 Bonn, Germany}

\author{M.~Aguena}
\affiliation{Laborat\'orio Interinstitucional de e-Astronomia - LIneA, Rua Gal. Jos\'e Cristino 77, Rio de Janeiro, RJ - 20921-400, Brazil}

\author{S.~Allam}
\affiliation{Fermi National Accelerator Laboratory, P. O. Box 500, Batavia, IL 60510, USA}
\author{S.~W.~Allen}

\author{O.~Alves}
\affiliation{Department of Physics, University of Michigan, Ann Arbor, MI 48109, USA}

\author{F.~Andrade-Oliveira}
\affiliation{Department of Physics, University of Michigan, Ann Arbor, MI 48109, USA}

\author{D.~Brooks}
\affiliation{Department of Physics \& Astronomy, University College London, Gower Street, London, WC1E 6BT, UK}

\author{A.~Carnero~Rosell}
\affiliation{Instituto de Astrofisica de Canarias, E-38205 La Laguna, Tenerife, Spain}
\affiliation{Laborat\'orio Interinstitucional de e-Astronomia - LIneA, Rua Gal. Jos\'e Cristino 77, Rio de Janeiro, RJ - 20921-400, Brazil}
\affiliation{Universidad de La Laguna, Dpto. Astrofísica, E-38206 La Laguna, Tenerife, Spain}

\author{M.~Carrasco~Kind}
\affiliation{Center for Astrophysical Surveys, National Center for Supercomputing Applications, 1205 West Clark St., Urbana, IL 61801, USA}

\author{J.~Carretero}
\affiliation{Institut de F\'{\i}sica d'Altes Energies (IFAE), The Barcelona Institute of Science and Technology, Campus UAB, 08193 Bellaterra (Barcelona), Spain}

\author{M.~Costanzi}
\affiliation{Astronomy Unit, Department of Physics, University of Trieste, via Tiepolo 11, I-34131 Trieste, Italy}
\affiliation{INAF-Osservatorio Astronomico di Trieste, via G. B. Tiepolo 11, I-34143 Trieste, Italy}
\affiliation{Institute for Fundamental Physics of the Universe, Via Beirut 2, 34014 Trieste, Italy}

\author{L.~N.~da Costa}
\affiliation{Laborat\'orio Interinstitucional de e-Astronomia - LIneA, Rua Gal. Jos\'e Cristino 77, Rio de Janeiro, RJ - 20921-400, Brazil}

\author{T.~M.~Davis}
\affiliation{School of Mathematics and Physics, University of Queensland,  Brisbane, QLD 4072, Australia}
\author{S.~Desai}
\affiliation{Department of Physics, IIT Hyderabad, Kandi, Telangana 502285, India}
\author{J.~De~Vicente}
\affiliation{Centro de Investigaciones Energ\'eticas, Medioambientales y Tecnol\'ogicas (CIEMAT), Madrid, Spain}

\author{H.~T.~Diehl}
\affiliation{Fermi National Accelerator Laboratory, P. O. Box 500, Batavia, IL 60510, USA}

\author{S.~Dodelson}
\affiliation{Department of Physics, Carnegie Mellon University, Pittsburgh, Pennsylvania 15312, USA}
\affiliation{NSF AI Planning Institute for Physics of the Future, Carnegie Mellon University, Pittsburgh, PA 15213, USA}

\author{P.~Doel}
\affiliation{Department of Physics \& Astronomy, University College London, Gower Street, London, WC1E 6BT, UK}
\author{S.~Everett}
\affiliation{California Institute of Technology, 1200 East California Blvd, MC 249-17, Pasadena, CA 91125, USA}

\author{B.~Flaugher}
\affiliation{Fermi National Accelerator Laboratory, P. O. Box 500, Batavia, IL 60510, USA}
\author{J.~Frieman}
\affiliation{Fermi National Accelerator Laboratory, P. O. Box 500, Batavia, IL 60510, USA}
\affiliation{Kavli Institute for Cosmological Physics, University of Chicago, 5640 South Ellis Avenue, Chicago, IL 60637, USA}

\author{J.~Garc\'ia-Bellido}
\affiliation{Instituto de Fisica Teorica UAM/CSIC, Universidad Autonoma de Madrid, 28049 Madrid, Spain}

\author{R.~Gassis}
\affiliation{Department of Physics, University of Cincinnati, Cincinnati, OH 45221, USA}

\author{G.~Giannini}
\affiliation{Institut de F\'{\i}sica d'Altes Energies (IFAE), The Barcelona Institute of Science and Technology, Campus UAB, 08193 Bellaterra (Barcelona), Spain}
\affiliation{Kavli Institute for Cosmological Physics, University of Chicago, 5640 South Ellis Avenue, Chicago, IL 60637, USA}

\author{D.~Grün}
\affiliation{University Observatory, Faculty of Physics, Ludwig-Maximilians-Universit\"at, 
Scheinerstr. 1, D-81679 München, Germany}
\author{G.~Gutierrez}
\affiliation{Fermi National Accelerator Laboratory, P. O. Box 500, Batavia, IL 60510, USA}

\author{S.~R.~Hinton}
\affiliation{School of Mathematics and Physics, University of Queensland,  Brisbane, QLD 4072, Australia}
\author{D.~L.~Hollowood}
\affiliation{Santa Cruz Institute for Particle Physics, Santa Cruz, CA 95064, USA}

\author{D.~J.~James}
\affiliation{Center for Astrophysics \textbar\ Harvard \& Smithsonian, 60 Garden Street, Cambridge, MA 02138, USA}

\author{K.~Kuehn}
\affiliation{Australian Astronomical Optics, Macquarie University, North Ryde, NSW 2113, Australia}
\affiliation{Lowell Observatory, 1400 Mars Hill Rd, Flagstaff, AZ 86001, USA}

\author{O.~Lahav}
\affiliation{Department of Physics \& Astronomy, University College London, Gower Street, London, WC1E 6BT, UK}

\author{S.~Lee}
\affiliation{Kavli Institute for Particle Astrophysics \& Cosmology, Stanford University, Stanford, CA 94305, USA}
\author{M.~Lima}
\affiliation{Departamento de F\'isica Matem\'atica, Instituto de F\'isica, Universidade de S\~ao Paulo, CP 66318, S\~ao Paulo, SP, 05314-970, Brazil}
\affiliation{Laborat\'orio Interinstitucional de e-Astronomia - LIneA, Rua Gal. Jos\'e Cristino 77, Rio de Janeiro, RJ - 20921-400, Brazil}

\author{G.~Mahler}
\affiliation{STAR Institute, Quartier Agora - All\'ee du six Ao\^ut, 19c B-4000 Li\`ege, Belgium}
\affiliation{Centre for Extragalactic Astronomy, Durham University, South Road, Durham DH1 3LE, UK}
\affiliation{Institute for Computational Cosmology, Durham University, South Road, Durham DH1 3LE, UK}

\author{J.~L.~Marshall}
\affiliation{George P. and Cynthia Woods Mitchell Institute for Fundamental Physics and Astronomy, and Department of Physics and Astronomy, Texas A\&M University, College Station, TX 77843,  USA}

\author{R.~Miquel}
\affiliation{Instituci\'o Catalana de Recerca i Estudis Avan\c{c}ats, E-08010 Barcelona, Spain}
\affiliation{Institut de F\'{\i}sica d'Altes Energies (IFAE), The Barcelona Institute of Science and Technology, Campus UAB, 08193 Bellaterra (Barcelona), Spain}
\author{J.~Myles}
\affiliation{Department of Astrophysical Sciences, Princeton University, Peyton Hall, Princeton, NJ 08544, USA}

\author{R.~L.~C.~Ogando}
\affiliation{Observat\'orio Nacional, Rua Gal. Jos\'e Cristino 77, Rio de Janeiro, RJ - 20921-400, Brazil}

\author{M.~E.~S.~Pereira}
\affiliation{Hamburger Sternwarte, Universit\"{a}t Hamburg, Gojenbergsweg 112, 21029 Hamburg, Germany}

\author{A.~Pieres}
\affiliation{Laborat\'orio Interinstitucional de e-Astronomia - LIneA, Rua Gal. Jos\'e Cristino 77, Rio de Janeiro, RJ - 20921-400, Brazil}
\affiliation{Observat\'orio Nacional, Rua Gal. Jos\'e Cristino 77, Rio de Janeiro, RJ - 20921-400, Brazil}

\author{A.~A.~Plazas~Malag\'on}
\affiliation{Kavli Institute for Particle Astrophysics \& Cosmology, Stanford University, Stanford, CA 94305, USA}
\affiliation{SLAC National Accelerator Laboratory, 2575 Sand Hill Road, Menlo Park, CA 94025, USA}

\author{A.~Porredon}
\affiliation{Centro de Investigaciones Energ\'eticas, Medioambientales y Tecnol\'ogicas (CIEMAT), Madrid, Spain}
\affiliation{Ruhr University Bochum, Faculty of Physics and Astronomy, Astronomical Institute, German Centre for Cosmological Lensing, 44780 Bochum, Germany}

\author{C.~L.~Reichardt}
\affiliation{School of Physics, University of Melbourne, Parkville, VIC 3010, Australia}

\author{A.~K.~Romer}
\affiliation{Department of Physics and Astronomy, Pevensey Building, University of Sussex, Brighton, BN1 9QH, UK}

\author{S.~Samuroff}
\affiliation{Department of Physics, Northeastern University, Boston, MA 02115, USA}
\affiliation{Institut de F\'{\i}sica d'Altes Energies (IFAE), The Barcelona Institute of Science and Technology, Campus UAB, 08193 Bellaterra (Barcelona), Spain}

\author{E.~Sanchez}
\affiliation{Centro de Investigaciones Energ\'eticas, Medioambientales y Tecnol\'ogicas (CIEMAT), Madrid, Spain}

\author{D.~Sanchez Cid}
\affiliation{Centro de Investigaciones Energ\'eticas, Medioambientales y Tecnol\'ogicas (CIEMAT), Madrid, Spain}

\author{I.~Sevilla-Noarbe}
\affiliation{Centro de Investigaciones Energ\'eticas, Medioambientales y Tecnol\'ogicas (CIEMAT), Madrid, Spain}

\author{M.~Schubnell}
\affiliation{Department of Physics, University of Michigan, Ann Arbor, MI 48109, USA}

\author{M.~Smith}
\affiliation{Physics Department, Lancaster University, Lancaster, LA1 4YB, UK}
\author{E.~Suchyta}
\affiliation{Computer Science and Mathematics Division, Oak Ridge National Laboratory, Oak Ridge, TN 37831, USA}
\author{M.~E.~C.~Swanson}
\affiliation{Center for Astrophysical Surveys, National Center for Supercomputing Applications, 1205 West Clark St., Urbana, IL 61801, USA}

\author{A.~Tamošiūnas}
\affiliation{Case Western Reserve University, 10900 Euclid Ave., Cleveland, OH 44106, USA}
\author{G.~Tarle}
\affiliation{Department of Physics, University of Michigan, Ann Arbor, MI 48109, USA}

\author{D.~L.~Tucker}
\affiliation{Fermi National Accelerator Laboratory, P. O. Box 500, Batavia, IL 60510, USA}

\author{V.~Vikram}
\affiliation{High-Energy Physics Division, Argonne National Laboratory, 9700 South Cass Avenue, Lemont, IL 60439, USA}

\author{N.~Weaverdyck}
\affiliation{Department of Astronomy, University of California, Berkeley,  501 Campbell Hall, Berkeley, CA 94720, USA}
\affiliation{Lawrence Berkeley National Laboratory, 1 Cyclotron Road, Berkeley, CA 94720, USA}
\author{J.~Weller}
\affiliation{Max Planck Institute for Extraterrestrial Physics, Gießenbachstr.~1, D-85748 Garching, Germany}
\affiliation{University Observatory, Faculty of Physics, Ludwig-Maximilians-Universit\"at, 
Scheinerstr. 1, D-81679 München, Germany}
\author{P.~Wiseman}
\affiliation{School of Physics and Astronomy, University of Southampton,  Southampton, SO17 1BJ, UK}

\collaboration{the SPT and DES Collaborations}
\noaffiliation


\begin{abstract} % arXiv has a maximum of 1920 characters

We use galaxy cluster abundance measurements from the South Pole Telescope (SPT) enhanced by Multi-Component Matched Filter (MCMF) confirmation and complemented with mass information obtained using weak-lensing data from Dark Energy Survey Year~3 (DES Y3) and targeted Hubble Space Telescope (HST) observations for probing deviations from the cold dark matter paradigm. Concretely, we consider a class of dark sector models featuring interactions between dark matter (DM) and a dark radiation (DR) component within the framework of the Effective Theory of Structure Formation (ETHOS). We focus on scenarios that lead to power suppression over a wide range of scales, and thus can be tested with data sensitive to large scales, as realized for example for DM$-$DR interactions following from an unbroken non-Abelian $SU(N)$ gauge theory (interaction rate with power-law index $n=0$ within the ETHOS parametrization). Cluster abundance measurements are mostly sensitive to the amount of DR interacting with DM, parameterized by the ratio of DR temperature to the cosmic microwave background (CMB) temperature, $\xi_{\rm DR}=T_{\rm DR}/T_{\rm CMB}$. We find an upper limit $\xi_{\rm DR}<17\%$ at $95\%$ credibility. When the cluster data are combined with Planck 2018 CMB data along with baryon acoustic oscillation (BAO) measurements we find $\xi_{\rm DR}<10\%$, corresponding to a limit on the abundance of interacting DR that is around three times tighter than that from CMB+BAO data alone. We also discuss the complementarity of weak lensing informed cluster abundance studies with probes sensitive to smaller scales, explore the impact on our analysis of massive neutrinos, and comment on a slight preference for the presence of a non-zero interacting DR abundance, 
%leaving room to address the $S_8$ tension. 
which enables a physical solution to the $S_8$ tension.
\end{abstract}

\maketitle

%\tableofcontents

%===========================================================
\section{Introduction}
\label{sec:introduction}
%===========================================================
The abundance of galaxy clusters has long been recognized as a powerful cosmological probe \cite{White1993ClusterS8,Haiman2001ApJ...553..545H}, and in recent years, cluster number counts have provided competitive constraints on cosmological parameters~\cite{Vikhlinlin2009ApJ...692.1060V,Bocquet2019SPTcosmo, Bocquet2024SPTcosmo, Ghirardini:2024yni}. Clusters are the largest collapsed structures in the Universe, and their abundance is particularly sensitive to the density of matter $\Omega_{\rm m}$, the amplitude of density fluctuations at large scales, usually parameterized by $\sigma_8$ at 8$h^{-1}\,{\rm Mpc}$, and the dark energy equation of state parameter $w$. This makes cluster number counts an important tool for testing models that have an impact on structure formation and deviate from the standard cosmological model (\LCDM).

Understanding the nature of dark matter (DM) is a central problem of cosmology. Besides the ability to interact gravitationally, the identity and properties of DM are still unknown. Observations of the cosmic microwave background anisotropies (CMB)~\cite{Planck:2018vyg} and galaxy clustering via baryon acoustic oscillations (BAO) \cite{BOSS:2016wmc, Reid:2015gra,eBOSS:2020yzd}, together with weak lensing information \cite{Heymans:2020gsg, DES:2021wwk, Kilo-DegreeSurvey:2023gfr,Kilo-DegreeSurvey:2023gfr}, can be used as precision probes of certain fundamental DM properties. Future CMB and large scale structure (LSS) surveys \cite{Laureijs2011, EuclidTheoryWorkingGroup:2012gxx,Euclid:2021icp,CMB-S4:2016ple,Abazajian:2019eic,LSST:2008ijt,Euclid:2024pwi} are expected to be sensitive to even small deviations from \LCDM. 
In that regard, the tension in the value of $S_8=\sigma_8 \sqrt{\Omega_{\rm m}/0.3}$, can serve as a reference for searches beyond the cold dark matter (CDM) scenario. For $\Lambda$CDM, a tension of $\sim 1.5-3 \,\sigma$, depending on the dataset considered, has been reported between the high values obtained by the Planck 2018 analysis of CMB anisotropies~\cite{Planck:2018vyg} and structure formation probes such as KiDS~\cite{Heymans:2020gsg}, DES~\cite{DES:2021wwk}, HSC~\cite{HSC:2018mrq}, SPT~\cite{Bocquet2019SPTcosmo}, eROSITA (eFEDS)~\cite{Chiu:2022qgb}, eBOSS~\cite{Chabanier:2018rga}, except eROSITA (eRASS1)~\cite{Ghirardini:2024yni} which reported a higher value than CMB. Even in the absence of any strong evidence of deviations from the CDM paradigm, quantifying the extent to which the behavior of DM may deviate from that of a cold and collisionless matter component is still highly informative for identifying viable DM theories.

Extensions of DM models beyond the CDM paradigm include for example DM self-interaction on galactic~\cite{Tulin:2017ara,Bullock:2017xww} or even cosmological scales~\cite{Egana-Ugrinovic:2021gnu,Bottaro:2024pcb} and decays to other types of particles, see e.g.,~\cite{Simon:2022ftd, Fuss:2022zyt,Bucko:2022kss}. In this work we focus on the commonly considered scenario of a secluded dark sector where DM (or a fraction of it) interacts with a relativistic component called dark radiation (DR)~\cite{vandenAarssen:2012vpm,Buen-Abad:2015ova, Lesgourgues:2015wza,Cyr-Racine:2015ihg,Vogelsberger:2015gpr, Chacko:2015noa, Chacko:2016kgg, Raveri:2017jto, Buen-Abad:2017gxg, Archidiacono:2017slj, Pan:2018zha, Archidiacono:2019wdp, Becker:2020hzj, Rubira:2022xhb, Hooper:2022byl}. In this case, structure formation can be (slightly) suppressed on scales $\gtrsim h^{-1}\,$Mpc, making these types of models relevant for addressing the $S_8$ tension. Interacting dark matter$-$dark radiation (IDM$-$DR) scenarios arise within many classes of theoretical models, and their cosmological effect can be described within the framework of the Effective Theory of Structure Formation (ETHOS)~\cite{Cyr-Racine:2015ihg,Vogelsberger:2015gpr}. A comparison between different models and their impact on structure formation can be found in~\cite{Euclid:2024pwi}.

A particularly straightforward and predictive example for a microscopic IDM$-$DR model is based on a dark sector featuring a weakly coupled, unbroken $SU(N)$ gauge symmetry~\cite{Buen-Abad:2015ova, Lesgourgues:2015wza, Buen-Abad:2017gxg, Rubira:2022xhb}. In this scenario, DM consists of a particle species with gauge interactions governed by this symmetry, and the dark gauge bosons constitute the DR component. Their characteristic self-coupling gives rise to an interaction between IDM and DR that leads to a moderate suppression of the matter power spectrum over a  wide range of scales~\cite{Buen-Abad:2015ova}, such that this setup can explain low $S_8$ values and  be tested with observations sensitive to relatively large scales~\cite{Buen-Abad:2017gxg,Archidiacono:2019wdp,Rubira:2022xhb,Mazoun:2023kid,Euclid:2024pwi}. 
Given that $SU(N)$ gauge theories underlie both the strong and weak force in Nature, it is a plausible question whether interactions of DM are governed by a similar mechanism. Note that, cosmologically, DM$-$DR interactions within a dark sector are analogous to baryon$-$photon interactions within the visible sector. The IDM$-$DR scenario thus probes a well-motivated, fundamental property of DM, being described by a framework that is conceptually closely related to the baryon$-$photon drag force underlying CMB acoustic oscillations. 

In previous analyses~\cite{Archidiacono:2019wdp,Rubira:2022xhb} this model has been explored using Planck measurements~\cite{Planck:2018vyg} of CMB anisotropies, as well as BAO and galaxy clustering data from BOSS DR12~\cite{BOSS:2016wmc}. Results indicate a preferred region in parameter space where $\sigma_8$ (or $S_8$) drops to lower values in the presence of DR. In a recent analysis~\cite{Mazoun:2023kid}, the sensitivity of galaxy cluster counts to this model was investigated, including a forecast for ongoing and future surveys like SPT-3G and CMB-S4 complemented with next-generation weak-lensing data, such as those expected from the Euclid and Rubin surveys. This analysis found that cluster counts are particularly sensitive to the DR energy density, or equivalently the value of the temperature ratio $\xi_{\rm DR}=T_{{\rm DR}}/T_{{\rm CMB}}$, and will be able to discriminate between $\LCDM$ and IDM$-$DR models. A forecast of the sensitivity of weak-lensing shear measurements by Euclid~\cite{Euclid:2024pwi} as well as for future 21cm observations~\cite{Plombat:2024kla} has been carried out within a slightly different setup of the same model.

In this work, we perform an analysis of the IDM$-$DR model based on the abundance of galaxy clusters detected via the thermal Sunyaev-Zeldovich effect (tSZE) in the South Pole Telescope (SPT)~\cite{Carlstrom_2011} survey data and confirmed using the MCMF algorithm \cite{Klein2018MCMF,Klein2019MARDY3} applied to optical/NIR followup data. This cluster abundance analysis is informed by weak lensing data from the Dark Energy Survey (DES) Y3~\cite{Flaugher_2015, DES:2016jjg, DES:2018gui} and targeted weak-lensing measurements from the Hubble Space Telescope (HST)~\cite{Schrabback:2020hxx}. The cluster sample comprises 1,005 clusters constructed from the combined SPT-SZ, SPTpol ECS, and SPTpol 500d surveys~\citep{SPT:2014wbo, SPT:2019hnt, SPT:2023via, SPT:2023tib}, in the redshift range $0.25-1.78$. Weak-lensing data provide information on cluster masses and allow for constraining observable$-$mass relations empirically. We employ weak-lensing data from DES Year 3 for 688 clusters with redshift $z<0.95$ and HST data for 39 clusters with a redshift range $0.6-1.7$. The same combination of datasets has recently been analyzed in~\cite{Bocquet2024SPTcosmo} and showed competitive constraints on \LCDM, the dark energy equation of state, and the sum of neutrino masses. In this work, we follow the same data analysis framework.

The structure of this article is as follows: in Sec.~\ref{sec:formalism}, we review the IDM$-$DR setup and summarize the SPT cluster and HST/DES weak-lensing datasets and analysis strategy in Sec.~\ref{sec:data} and Sec.~\ref{sec:analysis}, respectively. Our main results are discussed in Sec~\ref{sec:results}, and we conclude in Sec.~\ref{sec:conclusion}. The Appendices contain results for an extended model setup as well as posteriors including all parameters entering the joint cluster abundance and weak-lensing mass calibration analysis.

%===========================================================
\section{IDM$-$DR model setup}
\label{sec:formalism}
%===========================================================
Interacting dark sector models can be realized by various particle physics setups. Yet, structure formation on large scales is only sensitive to certain characteristics of DM. The ETHOS framework offers a systematic approach for describing these models in terms of a few relevant  parameters that encapsulate the impact on structure formation, and for mapping them to the underlying particle physics model properties~\cite{Cyr-Racine:2015ihg}. ETHOS thus allows for constraining a wide class of generic DM models using observational cosmological data in an efficient way. In this framework, one can differentiate between various models based on the temperature (and hence redshift) dependence of the effective interaction rate between DM and DR. Specifically, the rate relevant for the drag force generated by DM$-$DR interactions within the dark sector can be parameterized as  $\Gamma_{\rm IDM-DR}(z)\propto \sum_n a_n  (1+z)^{n+1}$. Here $n$ characterizes the dependence on redshift $z$, and $a_n$ is the absolute interaction strength of inverse length dimension. Both of these parameters encapsulate the properties of the microscopic model. In previous works~\cite{Archidiacono:2019wdp, Rubira:2022xhb}, analyses were carried out assuming for simplicity an interaction rate described by a single power-law, and considering the cases $n=0,2,4$, that are characteristic for various classes of particle models differing by the dependence of the DM$-$DR cross section on the momentum transfer. The case $n=0$ turns out to be phenomenologically most interesting as it gives rise to an interaction rate that features the same redshift-dependence as the Hubble rate during radiation domination. This means the ratio $\Gamma_{\rm IDM-DR}/\mathcal{H}$ (where $\mathcal{H}$ is the conformal Hubble rate) stays constant over an extended period of time, impacting a wide range of perturbation modes with different scales entering the horizon, leading to a rather gradual suppression of the matter power spectrum. This is in sharp contrast to $n>0$ for which a cutoff similar to warm dark matter scenarios is predicted~\cite{Buen-Abad:2015ova}. In contrast, the power suppression extending over a wide range of scales obtained from IDM$-$DR interactions with $n=0$ offers a potential solution to the $S_8$ tension~\cite{Buen-Abad:2017gxg,Archidiacono:2019wdp,Rubira:2022xhb,Mazoun:2023kid,Euclid:2024pwi}. Furthermore, the $n=0$ case is particularly well-motivated from a particle physics perspective. Concretely, the dark sector interaction described by a weakly coupled, unbroken non-Abelian $SU(N)$ gauge theory predicts an IDM$-$DR interaction strength described by $n=0$ within ETHOS. In what follows, we will describe its main properties and quantities of interest.

Following previous work, we allow for the possibility that only a fraction of the total DM population interacts with DR, given by  
    \be \label{eq:fractiondef}
f_{\rm IDM} \equiv \frac{\Omega_{\rm IDM}}{\Omega_{\rm IDM}+\Omega_{\rm CDM}}\,,
    \ee
with $\Omega_{\rm IDM}$ and $\Omega_{\rm CDM}$ being the IDM and CDM density parameters. Within the non-Abelian dark sector model, the IDM component is described by a Fermionic particle species that transforms under a non-trivial representation of $SU(N)$, assuming the fundamental representation for concreteness. The non-interacting part of the DM is provided by a particle species that transforms trivially under $SU(N)$. The DR consists of a thermal bath of massless $SU(N)$ gauge bosons. Their self-interactions, being a characteristic feature of a non-Abelian gauge group, imply that DR behaves as a {\it fluid} component (rather than as free-streaming radiation). In general, the dark sector is allowed to have a different temperature than the visible sector, with a ratio that remains constant throughout the cosmological epochs considered in this work, and given by
    \be\label{eq:xidef}
\xi_{\rm DR} \equiv \frac{T_{\rm DR}}{T_{\rm CMB}}\,.
    \ee
The two parameters $f_{\rm IDM}$ and $\xi_{\rm DR}$ are the relevant model parameters for our analysis in the following. For convenience, we also present various derived parameters, such as the DR density parameter 
\be 
\Omega_{\rm DR}  = (N^2-1) \, \xi_{\rm DR}^4\, \Omega_\gamma \,,
\ee
in terms of $\Omega_\gamma$, the photon density parameter today.
The contribution of DR to the effective number of relativistic species is given by
\begin{equation}
\begin{split}
 \Delta N_{\rm eff} =& \frac{\rho_{\rm DR}}{\rho_{\rm 1\nu}} =  \frac{8}{7}\left( \frac{11}{4}\right)^{4/3} (N^2-1) \,\xi_{\rm DR}^4  \,,
 \end{split}
 \label{eq:deltaN}
\end{equation}
with $\rho_{\rm DR}$ and $\rho_{\rm 1\nu}$ being, respectively, the  energy densities for DR and one massless neutrino family.

 The interaction between IDM and DR affects the evolution of density and velocity perturbations in a way that is analogous to the well-known baryon$-$photon interactions. The set of evolution equations for the density contrast $\delta_{\rm IDM}$ and velocity divergence $\theta_{\rm IDM}$ obtained within the ETHOS framework is given by
\begin{eqnarray}
\dot{\delta}_{\rm IDM} + \theta_{\rm IDM} - 3 \dot{\phi} &=& 0 \,,
\\
\dot{\theta}_{\rm IDM} - c_{\rm IDM}^2k^2\delta_{\rm IDM} && \nn\\
+ \mathcal{H}\theta_{\rm IDM} - k^2\psi &=& \Gamma_{\rm IDM-DR} \, \left( \theta_{\rm IDM} - \theta_{\rm DR}\right)\,,\label{eq:idm_eq}
\end{eqnarray}
where $k=|\textbf{k}|$ is the comoving wave number and $c_{\rm IDM}$ is the adiabatic IDM sound speed. The gravitational potentials $\phi$ and $\psi$ within conformal Newtonian gauge act as a source for perturbations for continuity and Euler's equations, as usual. Importantly, within ETHOS, the Euler equation features a drag term depending on the relative difference between $\theta_{\rm IDM}$ and the DR velocity divergence $\theta_{\rm DR}$. This interaction rate can be written as a power law in redshift as  
 \begin{eqnarray} \label{eq:gamma_dm_dr}
 \Gamma_{\rm IDM-DR} (z)  &=& - \frac{4}{3}(\Omega_{\rm DR}h^2)\, a_{\rm dark}(1+z)\,, \nonumber
 \end{eqnarray}
 with $a_{\rm dark}$ a free parameter for quantifying the interaction strength, that has units of ${\rm Mpc}^{-1}$. Here $h$ is the dimensionless Hubble constant.  Note that this redshift-dependence matches the general parametrization discussed above, with power-law index $n=0$.
The DR component is treated as a fluid, in line with the non-Abelian self-interaction of dark gauge bosons, and therefore we only take into account the first two moments of its distribution: the overdensity $\delta_{\rm DR}$ and the velocity divergence $\theta_{\rm DR}$, for which we can write
\begin{eqnarray}
\dot{\delta}_{\rm DR} + \frac{4}{3}\theta_{\rm DR} - 4 \dot{\phi} &=& 0 \,,
\\
\dot{\theta}_{\rm DR} - \frac{1}{4}k^2\delta_{\rm DR}&&\nn\\
 {} + k^2\sigma^2_{\rm DR}  - k^2\psi &=& \Gamma_{\rm DR-IDM}\,\left( \theta_{\rm DR} - \theta_{\rm IDM}\right) \,, \label{eq:theta_DR}
\end{eqnarray}
with shear stress $\sigma_{\rm DR}$ being set to zero in the fluid limit, and interaction rate $\Gamma_{\rm IDM-DR} =   \left( 4\rho_{\rm DR}/3 \rho_{\rm IDM}\right) \times \Gamma_{\rm DR-IDM}$ fixed by energy-momentum conservation with $\rho_{\rm IDM}$ and $ \rho_{\rm DR}$ being the IDM and DR energy densities.

While we stress that the setup presented here applies in principle to all IDM$-$DR dark sector models with interaction rate that has a redshift-dependence described by $n=0$ within ETHOS, we note that within the dark sector described by the $SU(N)$ gauge interaction the IDM$-$DR rate can be related to its fundamental parameters, which include the DM mass $m_{\chi}$ and the dark gauge coupling constant $g_d$~\cite{Rubira:2022xhb}, 
\begin{eqnarray}\label{eq:Gamma_IDM_DR_SUN}
&\Gamma_{\rm IDM-DR}(z) = -\frac{1}{1+z}\frac{\pi}{18}\frac{\alpha_d^2}{m_\chi}2(N^2-1) \times \\
&  \left\{ T_\text{DR}^2 \left[ \ln\alpha_d^{-1} + c_0 + c_1g_d + {\cal O}(g_d^2)\right] + {\cal O}\left(\frac{T_\text{DR}^4}{m_\chi^2}\right)\right\}\,, \nonumber
\end{eqnarray}
where $\alpha_d=g_d^2/(4\pi)$ is the dark fine structure constant, and
\begin{eqnarray}
	 c_0 &=& 1+\text{ln}\left(\frac{6}{2N+N_f}\right)+\text{ln}(4\pi)-24\,\text{ln}\,(A)\,, \\
	c_1 &=& \frac{3\sqrt{2N+N_f}}{4\pi}\sqrt{\frac{3}{2}}\,,
\end{eqnarray}
with the Glaisher-Kinkelin constant $A\simeq 1.28243$ and $N_f$ being the number of additional light Fermionic degrees
of freedom in the dark sector (with $N_f=0$ in the most minimal version of the model). Eq.~(\ref{eq:Gamma_IDM_DR_SUN}) takes into account the Debye-screening by the DR plasma (see \cite{Rubira:2022xhb} for a derivation).
Using this result, the interaction strength parameter $a_{\rm dark}$ can be mapped to the fundamental parameters, 
\bea\label{eq:adarkSUN}
  a_\text{dark} &=& \frac{\pi}{12}\frac{\alpha_d^2}{m_\chi}\frac{(N^2-1)^{\frac12}}{\xi_{\rm DR}^2}\frac{T_{\rm CMB,0}^2}{\Omega_\gamma h^2} \nn \\
  && \times \left[ \ln\alpha_d^{-1} + c_0 + c_1g_d + {\cal O}(g_d^2)\right]\nn\\
  && \hspace{-1.5cm} = 0.91\cdot 10^9 \text{Mpc}^{-1} \left(\frac{\alpha_d}{10^{-4}}\right)^2\left(\frac{100\,\text{GeV}}{m_\chi}\right)\left(\frac{0.1}{\xi_{\rm DR}}\right)^2  \nn\\
  && \hspace{-1.3cm} \times (N^2-1)^{\frac12}  \left[ \ln\alpha_d^{-1} -(1.34+\ln N) +0.413\sqrt{N} g_d\right]\,. \nn\\
\eea

In this work we focus on the tight coupling limit, where the value of $a_{\rm dark}$ is large enough to keep the IDM and DR tightly coupled. This  requires $a_{\rm dark}\gtrsim 10^6\, {\rm Mpc}^{-1}$ (see~\cite{Mazoun:2023kid} for more details). Using Eq.~\eqref{eq:adarkSUN}, we stress that the tight coupling limit is easily realized, requiring only a very weak lower bound on the dark fine-structure of $\alpha_d\gtrsim 10^{-6}$, for $m_{\chi}=100\,{\rm GeV}$ and $\xi_{\rm DR} = 0.1$ assuming an $SU(3)$ theory. To further constrain the value of $\alpha_d$, we use other properties of the model. In a non-Abelian $SU(N)$ gauge theory the interaction becomes stronger at low energies because of the running coupling, which can lead to confinement. To ensure that the cosmological evolution occurs within the unconfined phase, we require $\alpha_d\lesssim0.1$. Furthermore, the observed ellipticity of the gravitational potential of the galaxy NGC720 leads to a bound on the long-range interaction strength~\cite{Agrawal:2016quu}, being $\alpha_d\lesssim 5\cdot10^{-3}$  for $m_{\chi}=100\,{\rm GeV}$ and $N=3$. More discussion on both constraints can be found in~\cite{Rubira:2022xhb}. Finally, IDM self-interaction with a cross-section of order $1\,{\rm cm}^{2}/{\rm g}$ is relevant for small-scale puzzles, like the core-cusp problem (see~\cite{Tulin:2017ara} for a review and also \cite{vandenAarssen:2012vpm, Vogelsberger:2015gpr}). This can be achieved by a dark fine structure constant of order $\alpha_d\sim 10^{-5}$, depending on $m_\chi$. We note that this value is consistent with the lower bound required for the tight coupling limit as well as the upper bounds from confinement and galaxy ellipticity.

In summary, this IDM$-$DR setup can be characterized by three parameters: the fraction $f_{\rm IDM}$ of interacting DM, Eq.~(\ref{eq:fractiondef}), the amplitude $a_\text{dark}$ of this interaction, Eq.~(\ref{eq:adarkSUN}), and the temperature of dark radiation relative to the CMB $\xi_{\rm DR}$, Eq.~(\ref{eq:xidef}). Increasing $a_\text{dark}$ and $f_{\rm IDM}$ enhances the suppression of the matter power spectrum relative to \LCDM, but this has only a minor impact on the halo mass function within the regime probed by cluster counts, resulting in limited constraining power from this type of observation. In contrast, varying $\xi_{\rm DR}$ leads to a change of the scale on which the matter power spectrum is suppressed, having a substantial impact on the halo mass function, and consequently leading to a high sensitivity of galaxy cluster abundance measurements to this model parameter. For a more complete description of the effect of each parameter, we refer to~\cite{Mazoun:2023kid}. Throughout the main part of this work we consider the tightly coupled regime of large $a_\text{dark}$, and therefore in our analysis the two free model parameters, $\xi_{\rm DR}$ and $f_{\rm IDM}$, are sufficient to characterize the IDM$-$DR model.

\begin{figure*}
    \centering
    \includegraphics[width=0.47\textwidth]{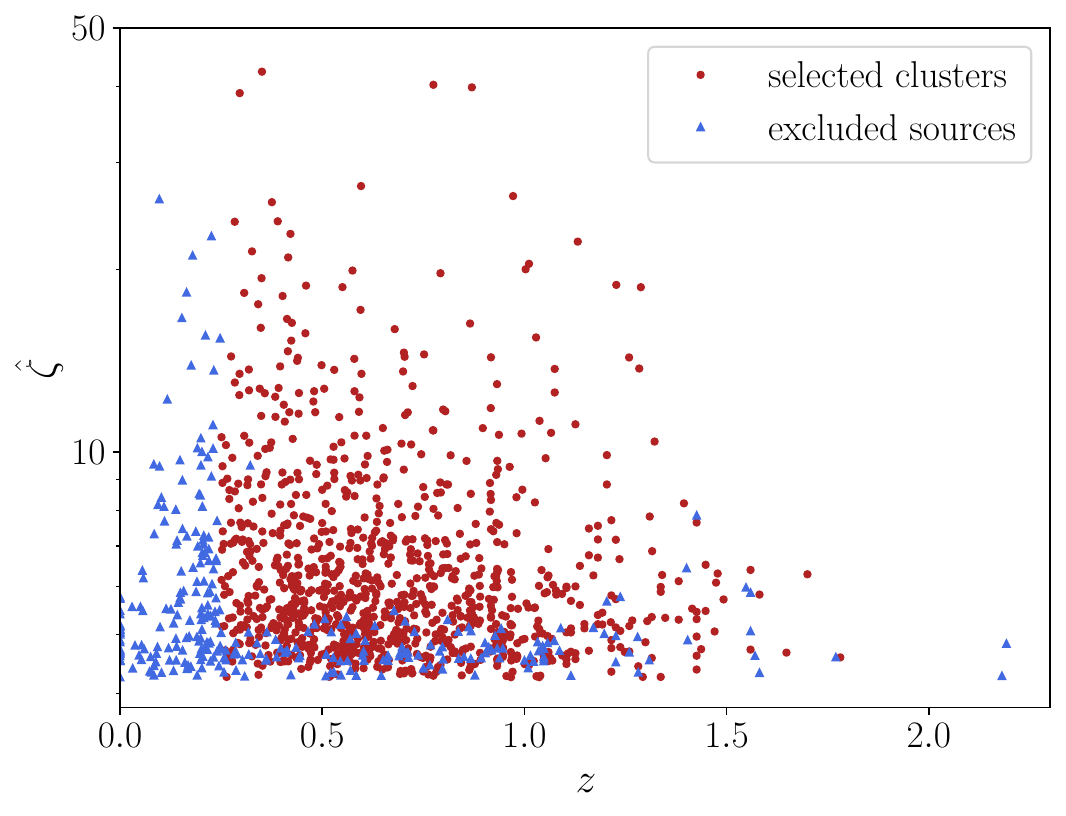}
    \includegraphics[width=0.48\textwidth]{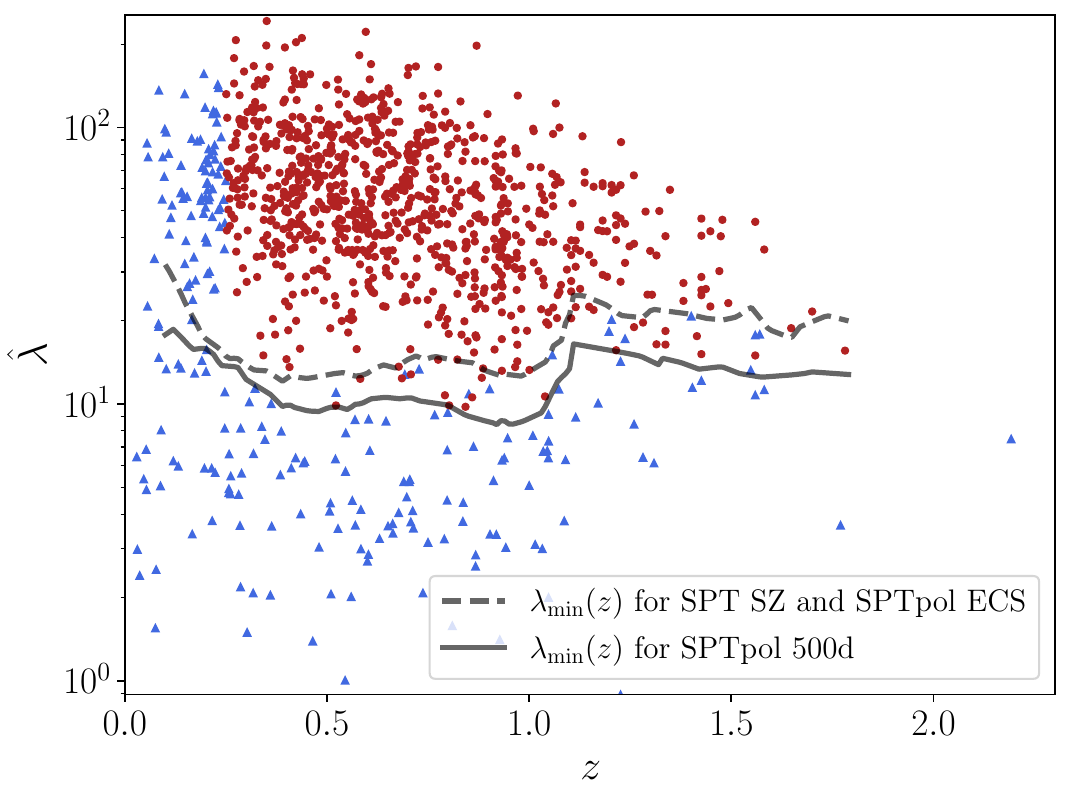}
    \caption{The distribution of clusters and contaminating sources from the SPT surveys \cite{SPT:2019hnt,SPT:2023via, SPT:2023tib} in the observables tSZE significance $\hat\zeta$, optical or NIR richness $\hat\lambda$ and redshift $z$. 
    %We show the excluded clusters ($z<0.25$) and contamination in blue, and the analysed sample of confirmed clusters in red.  
    The red points are confirmed clusters that  
    meet the selection criteria listed in Eq.\,\eqref{eq:select_DES} and constitute the sample analyzed here.  The blue points at low $\hat\zeta$ (left) and $\hat\lambda$ (right) consist of contaminants in the original tSZE-selected candidate list and confirmed clusters at $z<0.25$. The optical followup with the MCMF algorithm delivers cluster redshifts, excludes the contaminants and enhances the total number of confirmed clusters by $\sim$30\% in comparison to a purely tSZE selected sample with the same final contamination level of $\sim2$\%. 
    The figures illustrate that the bulk of the confirmed clusters lie well above the $\lambda_{\rm min}(z)$ MCMF selection threshold and that the bulk of the confirmed sample lies at redshifts $z<1$. Note that the feature in the $\lambda_{\rm min}(z)$ selection at $z=1.1$ corresponds to the inclusion of WISE data at higher redshift.
    }
    \label{fig:data_SNR-richness}
\end{figure*}
%===========================================================
\section{Cluster catalog and weak-lensing datasets}
\label{sec:data}
%===========================================================
We summarize the datasets used for the galaxy cluster abundance analysis in this work, covering first the MCMF enhanced SPT cluster catalog and then the DES Y3 and HST weak-lensing data, which are used for cluster mass calibration. Further details can be found in~\cite{Bocquet2023SPTmethod}. 

%-----------------------------------------------------------
\subsection{SPT cluster catalog}
\label{sec:SPT}
%-----------------------------------------------------------
The SPT cluster catalog is constructed from three multi-frequency mm-wave surveys: SPT-SZ, SPTpol ECS, and SPTpol 500d \citep{SPT:2014wbo, SPT:2019hnt, SPT:2023via, SPT:2023tib}. Together, the surveys cover 5,270~deg$^2$ of the southern sky. Within the SPT-SZ observed area, the SPTpol 500d patch was re-observed to a greater depth, and we use only data from the latter in this overlapping region. Cluster candidates are selected using the observed tSZE {\it detection significance} $\hat\zeta$, which is the maximum detection signal-to-noise ratio (SNR) when scanning over a range of cluster angular sizes.  Clusters are then confirmed using optical and NIR data, which delivers 1) the observed {\it richness}, $\hat\lambda$, corresponding to the weighted number of passive galaxies in a cluster, and 2) the cluster redshift $z$.  For the clusters from the SPT-SZ and SPTpol surveys, the final sample size was increased by about $\sim$50\% using the MCMF algorithm, which allows one to use the distribution of richness measured along random lines of sight to push to lower tSZE detection significance $\hat\zeta$, while maintaining high purity in the final, confirmed cluster sample \cite{Klein2018MCMF,Klein2024SPT-SZ}. This same technique has been employed to construct the largest all-sky X-ray selected cluster sample \cite{Klein2023RASS}, the deepest Planck tSZE selected sample \cite{Hernandez2023Planck} and the largest tSZE selected cluster sample to date, which is based on ACT observations \cite{Klein2024ACTDR5}.

The cluster catalog over the entire survey region includes only clusters at $z>0.25$. Clusters at lower redshift are excluded because the detection filter used for the SPT maps that removes atmospheric noise and noise contributions from the primary CMB measurement also removes a significant fraction of the signal from the lower redshift clusters.

The overlapping region between the SPT surveys and DES Y3 weak-lensing dataset covers 3,567~deg$^2$ ($\sim$70\% of the SPT survey area). The confirmation process with MCMF~\cite{Klein2018MCMF} leads to the cluster redshift $z$, a richness $\hat\lambda$, and the optical center of the cluster on the sky. A tSZE selected SPT cluster candidate is considered a confirmed cluster only if its richness exceeds a richness threshold $\lambda_{\rm min}(z)$. To ensure the same sample purity (of $\sim 98\%$) at all redshifts, the lower limit in richness varies with cluster redshift $\lambda_{\rm min} (z)$; more details are provided in the SPT catalog papers~\cite{Klein2024SPT-SZ,Klein2024OSPTpol}. MCMF was run using optical DES Y3 galaxies as well as using infrared WISE galaxies~\cite{Wright_2010}. At high redshifts ($z>1$)  the optical confirmation significantly suffers from the limited depth of DES Y3 data and is surpassed in sensitivity by the infrared confirmation at $z>1.1$.  

The various SPT surveys have different depths, resulting in cluster candidate lists with varying purity levels at a fixed detection significance, $\hat\zeta$. To maintain roughly constant purity in the cluster samples from the different SPT surveys, we apply different detection significance and richness thresholds. We select clusters based on the following criteria
\begin{equation}
  \label{eq:select_DES}
  \begin{split}
    \hat\zeta&>4.25 \,/\, 4.5 \,/\, 5 \,\,(\text{500d / SZ / ECS}), \\
    \hat\lambda&>\hat\lambda_\mathrm{min}(z), \\
    z&>0.25.
  \end{split}
\end{equation}
For the 1,327~deg$^2$ of the SPT survey not covered by DES Y3, we select samples by applying a cut in the SPT detection significance $\hat \zeta>5$. The resulting cluster candidate list has a purity of $\gtrsim95\%$. Cluster confirmation and redshift assignment are carried out using targeted optical observations (like PISCO; \cite{stalder14}). More details can be found in the original catalog publication \cite{SPT:2019hnt}.

The final sample consists of 1,005 confirmed clusters with tSZE significance, richness, and redshift measurements. Additionally, all clusters in the DES Y3 region have richness measurements and optically determined cluster centers provided with the MCMF algorithm operating on the DES Y3 or WISE datasets. We show the sample tSZE significance and richness as a function redshift in Fig.~\ref{fig:data_SNR-richness}.

\begin{figure}
    \centering
    \includegraphics[width=0.98\linewidth]{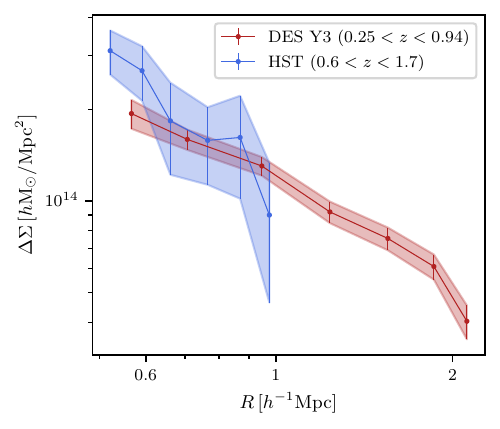}
    \caption{Average matter profiles $\Delta\Sigma(R)$ for MCMF confirmed SPT selected clusters as extracted from DES Y3 data for 688 clusters (red) and from HST data for 39 clusters (blue). Error bars correspond to the 1$\sigma$ region.}
    \label{fig:WL-data}
\end{figure}

%-----------------------------------------------------------
\subsection{DES Y3 weak-lensing data}
\label{sec:DES}
%-----------------------------------------------------------
The DES Y3 dataset includes three years of survey data.
It covers approximately 5,000~deg$^2$ of the southern sky. The data were collected using the 570-megapixel Dark Energy Camera (DECam) mounted on the Blanco 4-meter telescope at the Cerro Tololo Inter-American Observatory (CTIO) in Chile. The shape catalog was constructed using the {\scshape Metacalibration} method~\cite{Huff:2017qxu, Sheldon:2017szh}. Further information about systematics and calibration can be found in the DES Y3 catalog dedicated papers; \citep{DES:2020aks} for photometric dataset, \citep{DES:2020vau} for point-spread function modeling, and \citep{DES:2020lsz, DES:2020jnm} for image and survey simulations. After applying all source galaxy selection cuts, the DES Y3 shear catalog includes approximately 100~million galaxies over a solid angle of 4,143~deg$^2$ with an effective galaxy source density of 5-6 arcmin$^{-2}$.  

A total of 3,567~deg$^2$ of the DES Y3 dataset overlaps the SPT surveys. We select lensing source galaxies in four tomographic bins following the 3$\times$2~pt analysis~\cite{DES:2021wwk} with their redshift distribution calibrated using self-organizing maps~\citep{DES:2020ebm}. For every SPT cluster, we extract a weak-lensing shear profile within the radial range $500 \, h^{-1} \, \text{kpc} < R < \frac{3.2}{1 + z} \, h^{-1} \, \text{Mpc}$, centered on the MCMF determined optical cluster center.
%\footnote{This choice excludes the core region that might be impacted by IDM self-interaction.}.
We limit the use of DES Y3 weak-lensing data to clusters with a redshift below $z=0.95$, leading to 688 cluster shear profiles extracted using a sample of 555,912 source galaxies. We show in Fig.~\ref{fig:WL-data} the constructed averaged matter profiles for confirmed MCMF SPT clusters. The discussion of relevant sources of uncertainty including cluster member contamination, miscentering of the shear profile, shear and photo-$z$ calibration, halo mass modeling, and impact of large-scale structure can be found in the dedicated cluster cosmology methods paper for this combination of data~\cite{Bocquet2023SPTmethod}.
%-----------------------------------------------------------
\subsection{High-redshift HST weak-lensing data}
\label{sec:HST}
%----------------------------------------------------------- 
DES Y3 data loses its constraining power at higher redshift ($z\gtrsim 0.9$), and therefore, we include also HST weak-lensing data. Because of its high resolution imaging, it provides very precise targeted measurements of shear profiles for clusters at higher redshift. We use the previously analyzed weak-lensing measurements for 39 clusters in the redshift range $0.6-1.7$ in~\cite{SPT:2016gov, Raihan:2020qnd, Hernandez-Martin:2020ogo, Schrabback:2020hxx, Zohren:2022bdi}. For a further description of this component of the dataset, we refer the reader to these papers. The constructed averaged matter profiles are shown in Fig.~\ref{fig:WL-data}.

%===========================================================
\section{Abundance and mass calibration analyses}
\label{sec:analysis}
%===========================================================
In this section, we summarize the analysis method used for the abundance and mass calibration likelihoods. We introduce the observable$-$mass relations, describe the method used to obtain the mass information from weak-lensing measurements and then present the calibration of the weak-lensing mass to halo mass mean relation (\Mwl-\Mhalo), which we use to model systematic uncertainties in the weak-lensing mass calibration. Finally, we describe the ingredients entering the pipeline, including the relevant likelihoods and priors on parameters. Our description follows the presentation in~\cite{Bocquet2023SPTmethod}.

%-----------------------------------------------------------
\subsection{Observable Mass Relations}
\label{sec:analysis_method}
%-----------------------------------------------------------
The observed tSZE significance $\hat\zeta$ is related to an unbiased or intrinsic detection significance $\zeta$ as~\cite{Vanderlinde_2010}
\begin{equation} 
\label{eq:zeta_to_observed}
P(\hat\zeta|\zeta) = \mathcal N\left(\sqrt{\zeta^2 + 3}, 1\right)\,, 
\end{equation}
where ${\cal N}(a,\sigma)$ refers to a Gaussian distribution with mean $a$ and standard deviation $\sigma$. The normal distribution arises from Gaussian noise in the survey maps. The bias correction factor of 3 accounts for noise bias introduced during the matched-filter search for peaks, considering three parameters: two for the location on the sky and one for the effective core radius of the tSZE signature.

The observable$-$mass relation, which links the mean intrinsic significance $\zeta$ to the halo mass $M_{200c}$ used in describing the Halo Mass Function (HMF), is modeled as a power law in both mass and the redshift-dependent dimensionless Hubble parameter ($E(z)\equiv H(z)/H_0$) 
\begin{equation}
  \begin{split}
    \langle\ln\zeta\rangle =& \ln\asz + \bsz \ln\left(\frac{M_{200c}}{3\cdot 10^{14}\,h^{-1}\Msun}\right) \\
    & + \csz \ln\left(\frac{E(z)}{E(0.6)}\right)\,. \label{eq:zetaM}
  \end{split}
\end{equation}
The parameters $\asz$, $\bsz$, and $\csz$ represent the relation amplitude, mass trend, and redshift trend, respectively. The quantity $M_{200c}$ denotes the halo mass within a region where the overdensity is 200 times the critical density at the cluster redshift. The pivot redshift and mass for the power law relation are set to $z=0.6$ and $M_{200c}=3\cdot 10^{14}h^{-1}M_\odot$, respectively. Both observed and simulated clusters exhibit structural variations due to factors such as the time elapsed since their last major merger, which introduces scatter in the observable at any given mass and redshift. We model the intrinsic scatter of the intrinsic significance $\zeta$ around the mean relation as a log-normal distribution characterized by its mass and redshift independent RMS (root mean square) variation, $\sigma_{\ln\zeta}$.

As the various SPT surveys have different depths, which we characterize using the parameter $\gamma_{\rm field}$, we rescale $\asz$ for each field with 
\be
\zeta_{0,\mathrm{field}} = \gamma_{\rm field} \asz\,. 
\ee
This variation in depth values affects the redshift evolution parameter $\csz$ as well, therefore, we rescale it for each SPT survey. A more complete discussion of the survey information and the rescaling of parameters are listed in~\cite{Bocquet2023SPTmethod}.

The noise in the observed richness $\hat \lambda$ of a particular cluster is represented by a Poisson sampling of an intrinsic richness $\lambda$. For $\lambda>10$, it is valid to take a lognormal limit for the Poisson distribution, this leads to
        \begin{equation}
        \label{eq:richness_to_observed}
        P({\rm ln}\, \hat \lambda |{\rm ln}\, \lambda) = \mathcal{N} ( {\rm ln}\,\lambda,  1/\lambda )  \, .
    \end{equation}
As for detection significance, we use the same ansatz for modeling the intrinsic richness 
\begin{equation}
  \begin{split}
    \langle\ln\lambda\rangle =& \ln\alambda + \blambda \ln\left(\frac{M_{200c}}{3\cdot 10^{14}\,h^{-1}\Msun}\right) \\
     &+ \clambda \ln\left(\frac{1+z}{1.6}\right) \,. \label{eq:lambdaM}
  \end{split}
\end{equation}
The richness$-$mass relation parameters $\alambda$, $\blambda$, and $\clambda$ correspond to the amplitude, mass trend, and redshift trend, respectively. The log-normal scatter of the intrinsic richness at a given mass and redshift around the mean richness is modeled using its RMS variation \sigmalnlambda.

We utilize two distinct measurements of richness: DES Y3 data for clusters with redshift $z < 1.1$ and WISE data for clusters with redshift $z > 1.1$. Instead of aligning the two richness measurements within the overlapping redshift range, we establish separate observable$-$mass relations for each, and transition from DES-based to WISE-based richnesses at $z = 1.1$.

We calibrate the observable$-$mass relations in eqs.~\eqref{eq:zetaM} and \eqref{eq:lambdaM} with their intrinsic scatter empirically with weak-lensing data. A detailed description of the method is presented for DES Y3 elsewhere~\cite{Bocquet2023SPTmethod}, but we briefly describe the main steps here.

The weak lensing (WL) observable for each cluster is the reduced tangential shear $\vec g_\mathrm{t}(R)$, which consists of a  single shear profile that merges all tomographic bins of weak-lensing source galaxies. We use this observable to determine the weak-lensing halo mass ($M_{WL}$) by fitting the reduced shear profile to a Navarro-Frenk-White (NFW) profile~\cite{Navarro1997ApJ...490..493N} for the mass distribution within the radial range of $500 \, h^{-1} \, \text{kpc} < R < \frac{3.2}{1 + z} \, h^{-1} \, \text{Mpc}$.

The inferred weak lensing mass \Mwl\, is not an exact determination of the halo mass, because halos do not individually match perfectly to the NFW model. We account for differences between \Mwl\ and the halo mass by establishing a \Mwl$-$\Mhalo\ mean relation~\cite{Becker2011ApJ...740...25B,Dietrich2019MNRAS.483.2871D,Grandis:2021aad}
    \begin{equation}
        \label{eq:WL_mass_rel}
        \begin{split}
         &\left\langle \mathrm{ln} \left( \frac{M_{\mathrm{WL}}}{2 \cdot  10^{14}\, h^{-1} M_\odot}  \right)  \right\rangle = 
         \aWL \\
        &\hspace{2.5cm}+ 
        \bWL \mathrm{ln} \left( \frac{M_{200c}}{2 \cdot 10^{14}\, h^{-1} M_\odot} \right) \, ,
        \end{split}
    \end{equation}
considering that \Mhalo\ is $M_{200c}$. \aWL\ is the logarithmic mass bias at $M_{200c}=2 \cdot 10^{14}\, h^{-1} M_\odot$ and \bWL\ is the mass trend in this bias. The width of the lognormal scatter around the mean of \Mwl\, is described as
    \begin{equation}
    \begin{split}
        \label{eq:WL_mass_var}
      \ln\sigma^2_{\ln\mathrm{WL}} = & \asigmaWL \\
      & + \bsigmaWL  \ln\left( \frac{M_{200c}}{2 \cdot 10^{14}\, h^{-1} M_\odot} \right) \,.
    \end{split}
    \end{equation}
    
We generate synthetic cluster shear maps by applying the source redshift distribution, cluster miscentering, and cluster member contamination to halo mass maps from numerical simulations (see description below). These synthetic shear maps are then used to calibrate the free parameters of the \Mwl$-$\Mhalo\ relation \citep{Grandis:2021aad}. 

Another way of parameterizing a parameter $p$ with its standard deviation $\Delta p$ is by splitting it as
\begin{equation}
    \label{eq:WL_param_modeling}
     p = \mathcal{N}(\bar p, \Delta p) = \bar p + \Delta p\, \mathcal{N}(0, 1)\, ,
\end{equation}
 where $\bar p$ is the mean value and $\Delta p$ is the uncertainty of the corresponding parameter $p$ obtained from simulations. To accurately describe the redshift dependent uncertainty of the logarithmic mass bias \aWL, the scatter in this parameter, $\Delta \aWL(z)$, is modeled as a linear combination of two redshift-dependent functions
\begin{equation}
    \Delta \aWL(z) =  \Delta_1 \aWL(z) +  \Delta_2 \aWL(z).
\end{equation}
To establish a relationship between \Mwl\ and the ``gravity-only mass'' \Mhalo\, we use mass maps from full-physics hydrodynamical simulations paired with gravity-only simulations with identical initial conditions. This approach includes the effect of baryons while leveraging the more robust halo mass function predictions from gravity-only simulations such as \citep{Tinker:2008ff}, including emulators like \cite{McClintock:2018uyf, Nishimichi:2018etk, Bocquet:2020tes}. In practice, we apply this method to the Magneticum~\cite{Hirschmann:2013qfl, Teklu_2015, Beck:2015qva, dolag2017distribution} and the Illustris TNG simulations~\cite{Pillepich:2017fcc, Marinacci:2017wew, Springel:2017tpz, Nelson:2017cxy, Naiman_2018, Nelson:2018uso}. The resulting constraints on the \Mwl$-$\Mhalo\ relation show some variation, which we attribute to uncertainties in modeling baryonic effects. To account for this, we inflate the uncertainty on all parameters of the \Mwl$-$\Mhalo\ relation by the additional uncertainty listed in Table~2 of~\cite{Grandis:2021aad}, which, for the amplitude \aWL, amounts to 0.02 or 2\%. The overall level of systematic uncertainty ($\sim 2-10\%$, See Fig~10 in~\cite{Bocquet2023SPTmethod}) is smaller than the current statistical uncertainty in the lensing dataset, ensuring that our analysis is not overly dependent on the accuracy of the Magneticum or Illustris TNG simulations in replicating the real universe.

HST-39 data is a targeted weak-lensing dataset for clusters. Given the selection of lensing sources via conservative color cuts, cluster member
contamination was found to be negligible for this sample (e.g.~\cite{Schrabback:2020hxx}), which is why it is not modeled here. Miscentring has been accounted for this sample as part of the mass bias estimation, as described in~\cite{Schrabback:2020hxx, Sommer:2021zdy}, assuming isotropic miscentring (see~\cite{Sommer:2023bvx} regarding the limitation of this assumption). The HST \Mwl$-$\Mhalo\ mean relation has only the first term in eq.~\eqref{eq:WL_mass_rel} (\aWL) and its scatter (\asigmaWL)~\cite{SPT:2016gov}.  

In summary, the observable$-$mass relations ($\zeta-$mass-$z$ and $\lambda-$mass-$z$) along with the \Mwl$-$\Mhalo\ relation are power law relations as described in Eqs.~\eqref{eq:zetaM}, \eqref{eq:lambdaM}, and \eqref{eq:WL_mass_rel}, respectively. All have an intrinsic scatter which we model as log-normal. We establish a covariance matrix between all of the scatter parameters, and account for possible correlations among all pairs. We introduce the parameters $\rho_{\zeta,\rm WL}$, $\rho_\mathrm{\zeta,\lambda}$, and $\rho_{{\rm WL},\lambda}$ for significance$-$weak-lensing, significance$-$richness, and weak-lensing$-$richness correlations, respectively. 

%-----------------------------------------------------------
\subsection{Cluster Abundance Likelihood}
\label{sec:likelihood}
%-----------------------------------------------------------

We use Bayes' theorem to infer values of cosmological parameters $\vec p$ assuming a cluster population model. We model the likelihood of cluster population as a Poisson realization of the halo observable function (HOF). The log-likelihood is given by 
\begin{equation}
\begin{split}
    \ln & \mathcal  L(\vec p) =  \sum_i \ln\frac{\dif^4 N(\vec p)}{\mathop{\dif \hat\zeta} \mathop{\dif\hat\lambda} \mathop{\dif \vec g_\mathrm{t}} \mathop{\dif z}}
    \Big|_{\hat\zeta_i, \hat\lambda_i, g_{\mathrm{t},i}, z_i}\\
     &- \idotsint\mathop{\dif\hat\zeta} \mathop{\dif\hat\lambda} \mathop{\dif \vec g_\mathrm{t}} \mathop{\dif z} 
     \frac{\dif^4 N(\vec p)}{\mathop{\dif\hat\zeta} \mathop{\dif\hat\lambda} \mathop{\dif \vec g_\mathrm{t}} \mathop{\dif z}} \Theta_\mathrm{s}(\hat\zeta,\hat\lambda,z)
     +\mathrm{const.}
\end{split}  
\label{eq:likelihood_start}
\end{equation}
Here $\Theta_\mathrm{s}$ is the survey selection function and it is defined in terms of the lower limit thresholds we impose in the observables $\hat \zeta$, $\hat \lambda$, and $z$ as described in Sec.~\ref{sec:SPT}.
The lensing data are tangential shear profiles $\vec g_\mathrm{t}$. The differential cluster abundance in observable space is defined as
    \begin{equation} 
        \label{eq:HOF_with_4_obs}
        \begin{split}
        \frac{\dif^4 N(\vec p)}{\dif\hat\zeta \dif\hat\lambda \dif \vec g_\mathrm{t} \dif z } =
        & \int\dif \Omega_\mathrm{s}\iiiint\dif M\, \dif\zeta\, \dif\lambda\, \dif M_\mathrm{WL}\, \\
        & P(\vec g_\mathrm{t}|M_\mathrm{WL}, \vec p)
        P(\hat\zeta|\zeta)
        P(\hat\lambda|\lambda) \\
        & P(\zeta, \lambda, M_\mathrm{WL} |M,z,\vec p)\\ 
        & \frac{\dif^2 N (\vec p,M,z)}{\dif M \dif V} \frac{\dif^2 V (\vec p,z)}{\dif z \dif \Omega_\mathrm{s}} 
        \, ,
        \end{split}        
    \end{equation}  
where $\frac{\dif^2 N(\boldsymbol{p},M, z)}{\mathop{\dif M} \mathop{\dif V}}$ is the halo mass function, $\frac{\dif^2 V(\boldsymbol{p},z)}{\mathop{\dif z} \mathop{\dif\Omega_\mathrm{s}}}$ is the differential volume, and $\Omega_\mathrm{s}$ is the survey footprint. The relations between observed and intrinsic parameters, $P(\hat\zeta|\zeta)$ and $P(\hat\lambda|\lambda)$, are defined in Eqs.~\eqref{eq:zeta_to_observed} and \eqref{eq:richness_to_observed}. $P(\zeta, \lambda, M_\mathrm{WL} |M,z,\vec p)$ follows from Eqs.~\eqref{eq:zetaM}, \eqref{eq:lambdaM}, \eqref{eq:WL_mass_rel} and \eqref{eq:WL_mass_var}. Finally, the lensing likelihood  $P(\vec g_\mathrm{t}|M_\mathrm{WL}, \vec p)$ is defined as a product of independent Gaussian probabilities in each radial bin $i$ of the tangential reduced shear profile of a given cluster
    \begin{equation}
    \begin{split}
        \label{eq:lensing_likelihood}
        P(\boldsymbol g_\mathrm{t}|M_\mathrm{WL}, \boldsymbol p) = & \prod_i \left(\sqrt{2\pi}\Delta g_{\mathrm{t},i} \right)^{-1} \\
        & e^{-\frac12 \left(\frac{ g_{\mathrm{t},i} - g_{\mathrm{t},i}(M_\mathrm{WL}, \boldsymbol p)}{\Delta g_{\mathrm{t},i}}\right)^2} \, ,
    \end{split}
    \end{equation}
with $\Delta g_{\mathrm{t},i}$ as the shape noise. Again, more discussion can be found in the cosmological methods paper~\cite{Bocquet2023SPTmethod}.

\begin{table}
  \caption{Priors on fitting parameters in the analysis of the abundance of SPT clusters with DES~Y3 and HST weak-lensing data.
  The lensing model parameters are informed by priors derived from simulations, as detailed in Sec.~\ref{sec:analysis_method}, and inform the empirical calibration of the other observable$-$mass relations. $\mathcal{U}$ represents a flat distribution, and $\mathcal{N}(a,\sigma)$ refers to a Gaussian distribution with mean $a$ and standard deviation $\sigma$.
  \label{tab:parameters}}
  \begin{ruledtabular}
    \begin{tabular}{lll}
      Parameter & Description & Informative Prior\\
      \colrule
      \multicolumn{3}{l}{DES Y3 cluster lensing} \\
      $\Delta_1 \aWL(z)$ & bias uncertainty as $f(z)$ & $\mathcal N(0, 1)$ \\
      $\Delta_2 \aWL(z)$ & bias uncertainty as $f(z)$ & $\mathcal N(0, 1)$ \\
      $\bWL$ & mass trend of bias & $\mathcal N(1.029, 0.006)$ \\
      $\asigmaWL$ & normalization of scatter & $\mathcal N(0, 1)$ \\
      $\bsigmaWL$ & mass trend of scatter & $\mathcal N(-0.226, 0.040)$ \\
      \colrule
      \multicolumn{3}{l}{HST cluster lensing} \\
      $\aWL$ & amplitude of bias & $\mathcal N(0, 1)$ \\
      $\asigmaWL$ & amplitude of scatter & $\mathcal N(0, 1)$ \\
      \colrule
      \multicolumn{2}{l}{tSZE $\zeta$-mass-$z$ parameters} \\
      $\ln\asz$ & amplitude & $\mathcal U(0.3, 2.0)$ \\
      \bsz & mass trend & $\mathcal U (1.2, 2.0)$\\
      \csz & redshift trend & $\mathcal U (-1.0, 1.5)$\\
      \sigmalnzeta & intrinsic scatter & $\mathcal U (0.05, 0.5)$\\
      $\gamma_\mathrm{ECS}$ & depth of SPTpol ECS & $\mathcal U (0.9, 1.2)$ \\
      \colrule
      \multicolumn{3}{l}{DES $\lambda$-mass-$z$ parameters (used for $z<1.1$)} \\
      $\ln\alambda$ & amplitude & $\mathcal U (3.0, 4.0)$\\
      \blambda & mass trend & $\mathcal U (0.7, 1.5)$\\
      \clambda & redshift trend & $\mathcal U (-1.0, 0.8)$\\
      \sigmalnlambda & intrinsic scatter & $\mathcal U (0.05, 0.5)$\\
      \colrule
      \multicolumn{3}{l}{WISE $\lambda$-mass-$z$ parameters (used for $z>1.1$)} \\
      $\ln\alambda$ & amplitude & $\mathcal U (3.0, 5.0)$\\
      \blambda & mass trend & $\mathcal U (0.7, 1.5)$\\
      \clambda & redshift trend & $\mathcal U (-4.0, 0.8)$\\
      \sigmalnlambda & intrinsic scatter & $\mathcal U (0.05, 0.5)$\\
      \colrule
      \multicolumn{3}{l}{Correlation coefficients} \\
      $\rho_{\zeta,\rm WL}$ & tSZE $\zeta-$WL & $\mathcal U(-0.5, 0.5)$ \\
      $\rho_\mathrm{\zeta,\lambda}$ & tSZE $\zeta-\lambda$ & $\mathcal U(-0.5, 0.5)$ \\
      $\rho_{{\rm WL},\lambda}$ & WL$-\lambda$ & $\mathcal U(-0.5, 0.5)$ \\
      \colrule
      \multicolumn{3}{l}{Cosmology} \\
      $\xi_{\rm DR}$ & DR temperature ratio & $\mathcal U (0.001, 0.5)$ \\
      \Om\ & matter density parameter & $\mathcal U (0.1, 0.5)$ \\
      %\Onuhh\ & neutrino density & $\mathcal U(0, 0.00644)$ \\
      \Obhh & baryon density parameter & $\mathcal N(0.02236,0.00015)$ \\
      $h$ & Hubble parameter & $\mathcal N(0.7, 0.05)$ \\
      $\ln10^{10}A_s$ & amplitude of $P(k)$ & $\mathcal U (2.0, 5.0 )$ \\
      $n_s$ & scalar spectral index & $\mathcal N(0.9649, 0.0044)$ \\
    \end{tabular}
  \end{ruledtabular}
\end{table}
%--------------------------------------------------
\subsection{Pipeline}
\label{sec:pipeline}
%-----------------------------------------------------------
The likelihood is implemented as a Python module in the framework of {\bf CosmoSIS}\footnote{\url{https://cosmosis.readthedocs.io/}} \cite{Zuntz:2014csq}. The matter power spectrum for the IDM$-$DR model is computed with the Boltzmann solver \texttt{CLASS} \cite{Lesgourgues:2011re,Lesgourgues:2011rg}.
%more details on the likelihood and its implementation can be found in ~\cite{Bocquet2023SPTmethod}. 
We explore the parameter space using the nested sampler {\bf MultiNest}~\cite{Feroz:2008xx,Feroz:2007kg,Feroz:2013hea}. For the purpose of comparison we include an analysis with CMB anisotropies from Planck 2018~\cite{Planck:2018vyg} and BAO data from BOSS DR12~\cite{BOSS:2016wmc}. Note that an analysis of IDM$-$DR with Planck and BAO data was previously performed in~\cite{Archidiacono:2019wdp,Rubira:2022xhb}. In this work, we consider the following dataset combinations
\begin{itemize} 
    \item SPT-cluster$\times$WL:  a sample of 1,005 tSZE selected clusters from SPT surveys~\citep{SPT:2014wbo, SPT:2019hnt, SPT:2023via, SPT:2023tib} confirmed with the MCMF algorithm and complemented with weak-lensing mass information from DES Y3~\cite{Flaugher_2015, DES:2016jjg, DES:2018gui} and HST~\cite{Schrabback:2020hxx} as described in Sec.~\ref{sec:data}
    \item CMB+BAO: CMB data from Planck 2018 (TT, TE, EE + lowE)~\cite{Planck:2018vyg} and BAO measurements from BOSS DR12~\cite{BOSS:2016wmc} 
\end{itemize}

For SPT-clusters$\times$WL we employ the full log-likelihood containing both the cluster abundance and the weak lensing mass calibration simultaneously as given in Eq.~\eqref{eq:likelihood_start}. We list priors for all parameters in Table~\ref{tab:parameters}. As discussed in Sec.~\ref{sec:analysis_method}, we employ informative, Gaussian priors on the weak-lensing mass calibration parameters introduced in Eqs.~\eqref{eq:WL_mass_rel}-\eqref{eq:WL_mass_var}, for DES Y3 data and HST data separately as described in Sec.~\ref{sec:analysis_method}. The observable$-$mass parameters introduced in Eqs.~\eqref{eq:zeta_to_observed}-\eqref{eq:lambdaM} are fully constrained by the mass calibration likelihood. We apply flat and uninformative priors on them. As explained in Sec.~\ref{sec:analysis_method}, we consider two richness$-$mass relations for the datasets (DES Y3 and WISE) at two different redshift regimes. Therefore, we apply priors for them separately. We then define flat priors for the correlation coefficients in the covariance matrix as mentioned in the previous section with a range which ensures that the resulting correlation matrix remains non-singular across all parameter combinations. Finally, we apply priors for cosmological parameters. When performing an analysis for galaxy cluster abundance only, we define Gaussian priors on \Obhh\ and $n_s$ from Planck results and a wide prior on $h$ covering all current available constraints [$\mathcal N(0.7, 0.05)$]. This is because clusters are not sensitive to the density of baryons and the spectral index, and they do not provide sufficient constraining power on the Hubble parameter by themselves. When including CMB and BAO data in the analysis, we apply only flat priors on the previously mentioned parameters, and marginalize over the optical depth of reionization $\tau$ as well.

The upper bound in the flat prior for $\xi_{\rm DR}$ arises from Planck constraints on $\Delta N_{\rm eff}$ for non-interacting DR, translated to  $\xi_{\rm DR}$ via Eq.~\eqref{eq:deltaN} (see also \cite{Archidiacono:2019wdp,Rubira:2022xhb}). We note that the precise choice is irrelevant for our results, because the constraints on DR interacting with DM that we derive in this work are far below this upper limit. Moreover, following~\cite{Mazoun:2023kid}, we focus on the tight coupling limit, where the interaction intensity is set to a sufficiently high value that our results are independent of its exact value. Specifically, we use ${\rm log}_{10}[a_{\rm dark}/{\rm Mpc^{-1}}]=8$. We also fix the IDM fraction to $f_{\rm IDM}=10\%$ to simplify the analysis and avoid parameter projection effects. Note that the matter power spectrum and HMF are relatively insensitive to this precise value within the relevant mass regime of the cluster sample we study here; more discussion on this choice can be found in~\cite{Mazoun:2023kid}. A complementary analysis including both $a_{\rm dark}$ and $f_{\rm IDM}$  as free parameters is provided in Appendix~\ref{sec:a&f}.

%===========================================================
\section{Results}
\label{sec:results}
%===========================================================

We present constraints on cosmological parameters within the IDM$-$DR model from galaxy cluster abundance based on the SPT sample with weak-lensing informed masses from DES Y3 and HST as described in Sec.~\ref{sec:data}.
In addition, we discuss the interplay of  the IDM$-$DR model with massive neutrinos and provide an assessment of the scale-dependent sensitivity of cluster abundance measurements in comparison to other datasets. Our full results including also all observable$-$mass parameters entering the analysis are shown in Appendix~\ref{app:full_plot}. 

\begin{figure*}
\centering
         \includegraphics[width=0.49\textwidth]{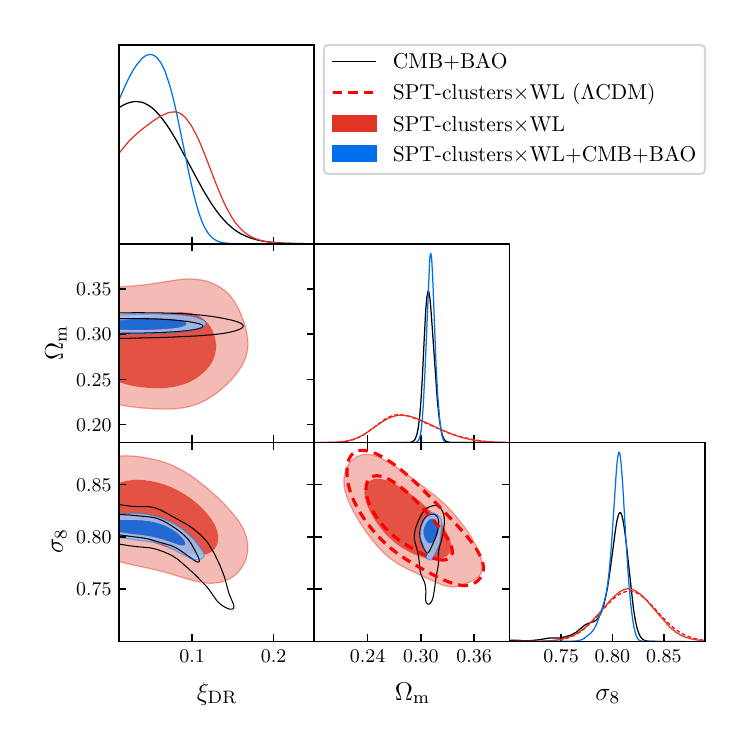}
         \raisebox{1.9cm}{$\includegraphics[width=0.49\textwidth]{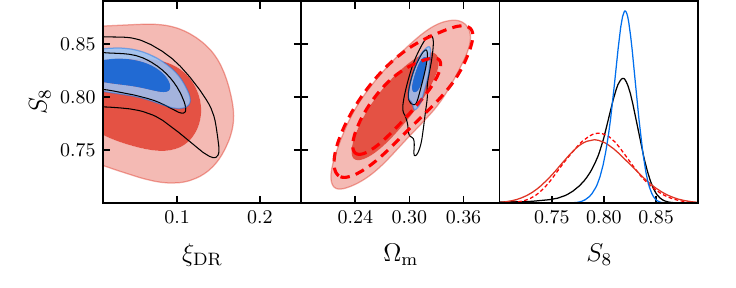}\atop
         \includegraphics[width=0.49\textwidth]{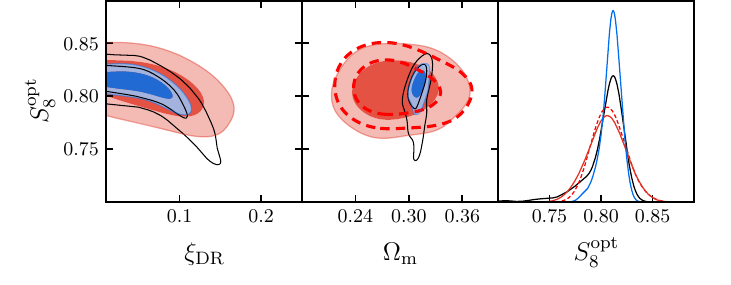}$}
 \caption{Posteriors obtained from analyzing the IDM$-$DR model with galaxy cluster abundance measurements (SPT tSZE selected and MCMF optical/NIR confirmed clusters from the SPT-SZ, SPTpol ECS, and SPTpol 500d surveys) combined with weak lensing data (DES Y3 and HST) for mass calibration (red). We show the results also in combination with CMB (Planck 2018) and BAO (BOSS DR12) measurements (blue), and compare with results for CMB+BAO alone (black lines). We include the analysis of cluster data within the \LCDM\ model from~\cite{Bocquet2024SPTcosmo} (dashed red lines) for comparison. On the left, constraints on $\xi_{\rm DR}$, \Om\ and \sig\ at 68\% credible intervals (CI) and 95\% upper limits are shown. On the right, we display constraints for the commonly used parameter combination $S_8=\sig (\Om/0.3)^{0.5}$, and $\Sopt=\sig (\Om/0.3)^{0.2}$ because it is the combination galaxy cluster counts are most sensitive to.}
 \label{fig:results}
\end{figure*}

\begin{table*}
  \caption{Parameter constraints for the IDM$-$DR and \LCDM\ models (marginal mean at 68\% CI, or upper limit at 95\% CI). $\Sopt=\sig (\Om/0.3)^{0.2}$ corresponds to the  combination that cluster counts are most sensitive to. We show results from SPT-clusters$\times$WL, CMB+BAO, and their combination. We compare to \LCDM\ results from SPT-clusters$\times$WL~\cite{Bocquet2024SPTcosmo}.
 \label{tab:results}}
  \begin{ruledtabular}
    \begin{tabular}{lcccccccc}
      Dataset & $\xi_{\rm DR}$ & $\Delta N_{\rm eff}$ \footnote{Note that here $\Delta N_{\rm eff}$ refers to DR that is tightly coupled to IDM and acts as a fluid. This explains the difference to the usually considered case of free-streaming, non-interacting extra radiation.} & \Om & \sig & $S_8$ &  \Sopt\  \\
      \colrule
      IDM$-$DR model\\
      \hskip0.5cm CMB+BAO & $< 0.130$ & $<0.010$ & $0.312\pm 0.005$ & $0.801\pm 0.019$ & $0.817\pm 0.021$ & $0.809\pm0.020$ \\
      \hskip0.5cm SPT-clusters$\times$WL & $< 0.166$ & $< 0.027$ & $0.285\pm 0.032$ & $0.815\pm 0.025$ & $0.793\pm 0.032$ & $0.805\pm0.018$\\
      %$0.803\pm0.019$ \\
      \hskip0.5cm SPT-clusters$\times$WL+CMB+BAO & $<0.098$ & $<0.003$ & $0.311 \pm 0.005$ & $0.804\pm 0.009$ & $0.819\pm 0.011$ & $0.810\pm 0.009$ \\
      %$0.812\pm 0.009$ \\
      \colrule
      \LCDM \, model\\
      \hskip0.5cm SPT-clusters$\times$WL & \dots & \dots & $0.286\pm0.032$ & $0.817\pm0.026$ & $0.795\pm0.029$ & $0.807\pm0.016$ \\
      %SPTclusters+lensing+{\it Planck} & \dots & $0.848\pm0.027$ & $0.783\pm0.026$ & $0.814\pm0.016$ & $0.75\pm0.04$ & $<0.6$ \\
    \end{tabular}
  \end{ruledtabular}
\end{table*}
%-----------------------------------------------------------
\subsection{Cosmological constraints in the IDM$-$DR model}
\label{sec:IDM$-$DR}
%-----------------------------------------------------------
We show our main results in Fig.~\ref{fig:results}\footnote{All plots are generated using GetDist \url{https://getdist.readthedocs.io/en/latest/}.} and summarize constraints on model parameters in Table~\ref{tab:results}. On the left side of Fig.~\ref{fig:results}, we present marginalized two-dimensional posteriors for \sig, \Om\, and $\xi_{\rm DR}$. In IDM$-$DR, SPT-clusters$\times$WL prefer slightly decreasing values of \sig\ when increasing $\xi_{\rm DR}$. A similar tendency was also pointed out when considering CMB and galaxy clustering data in~\cite{Archidiacono:2019wdp,Rubira:2022xhb}. This can be associated to a suppression of the fluctuation amplitude due to the IDM$-$DR interaction, that shifts to larger scales with increasing $\xi_{\rm DR}$. From SPT-clusters$\times$WL we obtain an upper bound on the DR temperature ratio $\xi_{\rm DR} <0.166$ at 95\% credibility. Adding clusters to CMB and BAO measurements yields a significant improvement in the joint constraint on the temperature ratio with $\xi_{\rm DR}<0.098$ (95\% credibility), compared to CMB+BAO alone $\xi_{\rm DR}<0.130$ (95\% credibility).

The upper bound on $\xi_{\rm DR}$ translates into an upper bound on $\Delta N_{\rm eff}$, as defined in Eq.~\eqref{eq:deltaN}. In an $SU(3)$ theory, this leads to $\Delta N_{\rm eff}<0.003$, which is about three times better than the constraint from CMB+BAO alone. Note that this constraint applies only to DR tightly coupled to IDM, and in the fluid limit. For reference, we note that this bound is $\sim 100$ times stronger than the one on free-streaming and non-interacting extra radiation from Planck CMB data combined with BAO ($\Delta N_{\rm eff}<0.28$~\cite{Abazajian:2019eic}). This can be explained by the strong impact of the DR density on the HMF due to its tight coupling to IDM and the associated effect on the matter power spectrum, which is absent for non-interacting and free-streaming DR.

As can be seen in Table~\ref{tab:results}, constraints on the parameters \Om\ and \sig\ from SPT-clusters$\times$WL within IDM$-$DR are broadly consistent with those obtained assuming \LCDM~\cite{Bocquet2024SPTcosmo}. On the right side of Fig.~\ref{fig:results}, we show the constraints on $S_8$ and $S_8^{\rm opt}$. In the light of the $S_8$ tension, we note that our results are consistent with both early time measurements (e.g. from Planck~\cite{Planck:2018vyg}) and late time measurements (e.g. cosmic shear measurements in the joint analysis of KiDS and DES data~\cite{Kilo-DegreeSurvey:2023gfr}) within the uncertainty range. Notably, when combining cluster data with CMB and BAO measurements, the constraining power on $S_8$ is two times improved compared to CMB+BAO only. Note that cluster abundance measurements are most sensitive to the parameter combination $S_8^{\rm opt}=\sig (\Om/0.3)^{0.2}$. As shown in the lower right part of Fig.~\ref{fig:results}, the degeneracy between \Sopt\ and \Om\ is completely broken. From  Table~\ref{tab:results}, we see that the uncertainty on  $S_8^{\rm opt}$ is considerably reduced by about $40\%$ compared to $S_8$ when considering SPT-clusters$\times$WL within IDM$-$DR. Moreover, the error on both $S_8^{\rm opt}$ as well as $S_8$ is only degraded by about $10\%$ within IDM$-$DR as compared to \LCDM. We further highlight the finding that the posteriors for \sig\, $S_8^{\rm opt}$ and $S_8$ feature a tail towards lower values, see Fig.~\ref{fig:results}, leaving room for addressing the $S_8$ tension within IDM$-$DR.

Interestingly, we also find a slight preference for a non-zero value of $\xi_{\rm DR}$. While not being statistically significant, we checked that this hint is robust to several variations in the analysis, such as including neutrino masses as in Sec.~\ref{sec:neutrinos}. We also note that it persists when including additionally $a_{\rm dark}$ and $f_{\rm IDM}$ as free model parameters, see Appendix~\ref{sec:a&f}. This is intriguing since both of these parameters become irrelevant in the limit $\xi_{\rm DR} \to 0$, leading to an increased volume of parameter space entering the marginalization, and therefore potentially a preference of the posterior for smaller values of $\xi_{\rm DR}$. Nevertheless, the hint for non-zero $\xi_{\rm DR}$ is also present in this case. It will be interesting to follow up on this hint with future cluster abundance measurements and complementary datasets.

\begin{figure}
    \begin{center}
    \includegraphics[width=0.98\columnwidth]{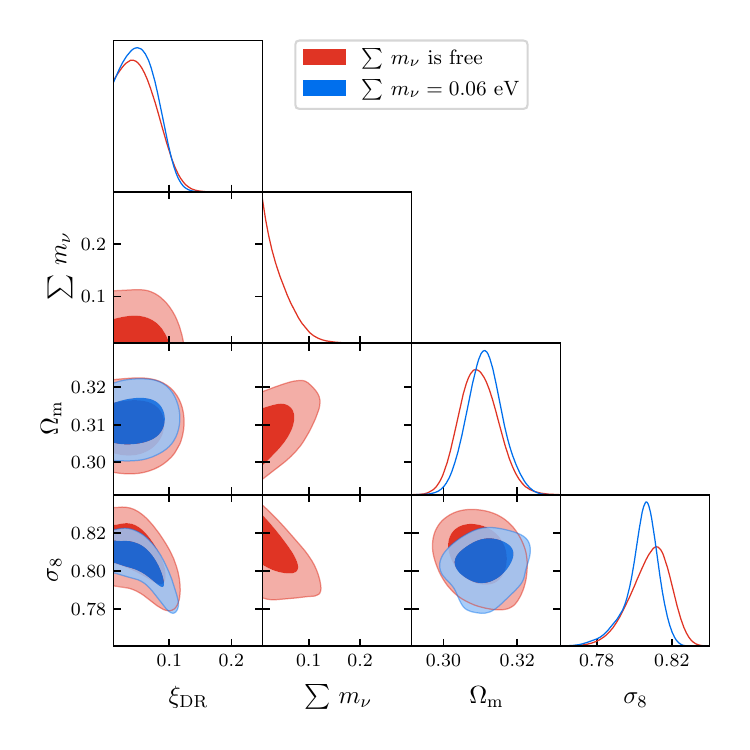}
    \caption{Comparison between the combined posteriors obtained when fixing the sum of neutrino masses to $\sum m_{\nu}=0.06\,{\rm eV}$ (blue), and when marginalizing over it (red). We show results for the combination SPT-clusters$\times$WL+CMB+BAO.}
    \label{fig:nufIDM}
    \end{center}
\end{figure}
%-----------------------------------------------------------
\subsection{Effect of neutrinos}
\label{sec:neutrinos}
%-----------------------------------------------------------
Massive neutrinos suppress structure formation below their free-streaming scale. As previously discussed in~\cite{Mazoun:2023kid}, this suppression is distinct from the one within the IDM$-$DR model, occurring at a different length scale. Additionally, neutrino masses also affect the background evolution after recombination, in contrast to IDM$-$DR. Although the effect of massive neutrinos is distinct from the one of the IDM$-$DR model, we carry out an analysis when adding the sum of neutrino masses as a free parameter, in order to check whether including it affects the constraint on $\xi_{\rm DR}$. 

We present a comparison between the case of fixing the sum of neutrino masses to $\sum m_{\nu}=0.06\,{\rm eV}$ \footnote{This is the minimum value for neutrino masses in the normal hierarchy as indicated by neutrino flavor oscillation experiments, for a review see~\cite{Lesgourgues:2006nd}.} and the case of letting it free in Fig.~\ref{fig:nufIDM}. We find that the sum of neutrino masses does not affect constraints on $\xi_{\rm DR}$ significantly. Importantly, there is no degeneracy between $\sum m_{\nu}$ and $\xi_{\rm DR}$. This agrees with the expectation from the sensitivity forecast of cluster abundance measurements from CMB-S4 and SPT-3G in~\cite{Mazoun:2023kid}. In turn, we also find that the neutrino mass bounds are not significantly degraded within IDM$-$DR as compared to $\Lambda$CDM. We obtain an upper bound $\sum m_{\nu}<0.096\,{\rm eV}$ when considering the data combination SPT-clusters$\times$WL+CMB+BAO. This upper limit is more stringent compared to $\sum m_{\nu}<0.13\,{\rm eV}$ obtained from CMB+BAO~\cite{Planck:2018vyg}.

\begin{figure}
    \begin{center}
    \includegraphics[width=0.98\columnwidth]{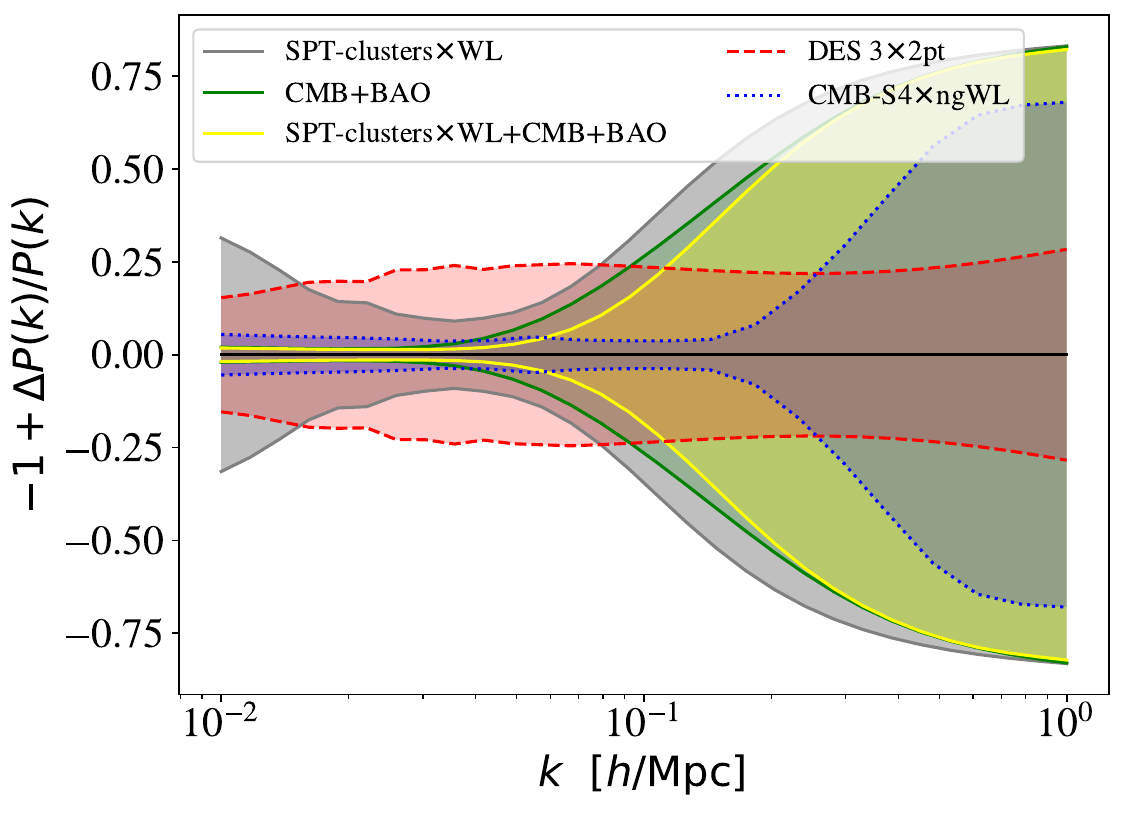}
    \caption{Variance ($1\sigma$ region) of the matter power spectrum as a function of wave number $k$ for different cosmological probes, inferred from  converged samples of model parameters. We set $\Delta P(k)=P(k)\pm \text{stdev}$, where $P(k)$ is the mean value of the matter power spectrum. Apart from SPT-clusters$\times$WL, CMB+BAO and their combination, we also show corresponding results from weak lensing shear measurements (DES $3\times 2 $pt results computed in the \LCDM\ model from~\cite{DES:2021wwk}) as well as a forecast for future cluster abundance measurements combined with next-generation weak-lensing data (CMB-S4$\times$ngWL, see~\cite{Mazoun:2023kid} for details). }
    \label{fig:pk}
    \end{center}
\end{figure}
%-----------------------------------------------------------
\subsection{Assessing the scale sensitivity and comparison to other probes}
\label{sec:pk}
%-----------------------------------------------------------
When testing for potential deviations from the \LCDM\ paradigm it is instructive to quantify to which range of length-scales a given measurement is dominantly sensitive to, and to which level of precision. Here we address this question  for cluster number counts, CMB+BAO as well as weak-lensing shear data and galaxy clustering. In Fig.~\ref{fig:pk} we show an approximate assessment of the scale sensitivity of SPT-clusters$\times$WL, CMB+BAO as well as their combination. We determine the linear matter power spectrum for each set of model parameters contained in the converged samples within our analysis based on the IDM$-$DR framework. We then display the variation of the power spectrum around the mean value within $1\sigma$ for each wave number, with $\Delta P(k)=(P(k)_{\rm mean}\pm \text{stdev})/P(k)_{\rm mean}$. 

As expected, we find that cluster abundance measurements are most sensitive to larger scales, approximately $k\in [0.02, 0.08]\,h/{\rm Mpc}$. This means they are less affected by the non-linear evolution of the universe and small scale effects like baryonic feedback. 
CMB+BAO data are highly sensitive to very large scales with small uncertainty. We note that the sensitivity of CMB+BAO measurements improves  when combining them with galaxy cluster data (as shown in yellow in Fig.~\ref{fig:pk}) in particular on scales $k\sim 0.1\,h/$Mpc that are particularly relevant for probing IDM$-$DR. All of these datasets loose sensitivity when going to smaller scales, in this case $k\gtrsim 0.1\, h/{\rm Mpc}$. This explains why clusters and CMB+BAO are able to constrain the value of $\xi_{\rm DR}$, which, as discussed and shown in Fig.~1 in~\cite{Mazoun:2023kid}, is responsible for defining the scale at which the suppression in the matter power spectrum sets in. The higher the value of $\xi_{\rm DR}$, the more the suppression shifts towards larger scales. This scale-dependence also explains the relatively weak sensitivity of both clusters as well as CMB+BAO data to the IDM$-$DR model parameters $a_{\rm dark}$ and $f_{\rm IDM}$. Constraining them would require including datasets that are more sensitive to smaller scales, such as  cosmic shear measurements, galaxy clustering, and their correlation (see DES Y3 $3\times2$pt results within $\Lambda$CDM in Fig.~\ref{fig:pk}). A combined analysis of cluster abundance with cosmic shear is thus a promising avenue for covering the entire IDM$-$DR model parameter space, with clusters providing sensitivity to the DR temperature $\xi_{\rm DR}$ and $3\times2$pt functions to the fraction of interacting DM $f_{\rm IDM}$ and/or the interaction strength $a_{\rm dark}$. We also note the sensitivity of cluster data will improve significantly with ongoing and future surveys. This is shown in blue in Fig.~\ref{fig:pk} for the case of clusters that could be detected with a CMB-S4 survey combined with next-generation weak-lensing data from Euclid or Rubin. This is owed to the larger size of the sample which will have more than an order of magnitude compared to current samples, and to the improved systematical uncertainties, more discussion can be found in~\cite{Mazoun:2023kid}. 

%===========================================================
\section{Conclusion}
\label{sec:conclusion}
%===========================================================
In this work, we derive constraints on interactions of dark matter with dark radiation using the tSZE detected SPT cluster sample consisting of $1,005$ clusters, with weak lensing measurements for 688 clusters from DES Y3 and for 39 clusters from HST, and employing the galaxy cluster abundance analysis framework developed in~\cite{Bocquet2023SPTmethod, Bocquet2024SPTcosmo}.
The cosmological IDM$-$DR model we consider has been discussed as a potential solution to the $S_8$ tension, and is part of the ETHOS framework that can describe a wide generic class of underlying particle physics models. A prominent example is a dark sector featuring a weakly coupled, unbroken non-Abelian $SU(N)$ gauge symmetry. The model is characterized by three parameters, being the temperature ratio $\xi_{\rm DR}$, the fraction of interacting dark matter $f_{\rm IDM}$, and the IDM$-$DR interaction strength $a_{\rm dark}$. The latter is related to the dark fine structure constant within the microscopic realization provided by the $SU(N)$ model. Galaxy cluster abundance data are mainly sensitive to $\xi_{\rm DR}$.

We find that the tSZE cluster sample obtained by SPT complemented with mass information from DES Y3 and HST yields an upper bound on $\xi_{\rm DR}< 17\%$ (95\% credibility). When combining these cluster abundance data with CMB data from Planck 2018 and BAO measurements from BOSS DR12, we obtain the most stringent to date upper bound $\xi_{\rm DR}< 10\%$ (95\% credibility).  This bound is $30\%$ tighter than the one from CMB+BAO alone. Note that this translates into an improvement in the upper bound on the DR energy density by a factor of three when adding clusters to CMB+BAO data.

Within the IDM$-$DR model, the cluster data set considered in this work yields $S_8=0.793\pm 0.032$ (68\% CI) as well as $\Sopt=\sig (\Om/0.3)^{0.2}=0.805\pm 0.018$  for the combination clusters are most sensitive to. This is consistent both with Planck CMB as well as with weak lensing shear measurements. Furthermore, slightly lower values of $S_8$ can be accommodated within the IDM$-$DR model compared to \LCDM, leaving room for addressing the $S_8$ tension within this class of models. Interestingly, we find a slight preference for a non-zero value of $\xi_{\rm DR}$, that is however statistically not significant. Yet, it occurs both when restricting the model to the limit for which DR and IDM are tightly coupled, as well as in the general case. 

Future galaxy cluster abundance data such as from CMB-S4 combined with mass information from Euclid or Rubin weak-lensing measurements will significantly increase the sensitivity to $\xi_{\rm DR}$. Furthermore, a combined analysis with weak lensing shear data could provide enhanced sensitivity to the model parameters $f_{\rm IDM}$ and $a_{\rm dark}$, and would be instrumental for scrutinizing the question of whether IDM$-$DR models can address the potential $S_8$ tension.

In summary, our results show that weak-lensing informed galaxy cluster abundance measurements are sensitive to fundamental properties of DM, testing its cold and collisionless nature with unprecedented sensitivity, and in a way that is complementary to both CMB and cosmic shear measurements.

\vspace*{2em}
\acknowledgments
 We acknowledge support by the Excellence Cluster ORIGINS, which is funded by the Deutsche Forschungsgemeinschaft (DFG, German Research
Foundation) under Germany’s Excellence Strategy - EXC-2094 - 390783311. We acknowledge support from the Max Planck Society Faculty Fellowship program at MPE, the Ludwig-Maximilians-Universit\"at in Munich, and the Technical University of Munich. The MCMC analysis was carried out at the Computational Center for Particle and Astrophysics (C2PAP) which is a computing facility from ORIGINS. Asmaa Mazoun thanks the mentoring program of ORIGINS, especially, Amelia Bayo Aran for her valuable advice.

The Innsbruck authors acknowledge support provided by the Austrian Research
Promotion Agency (FFG) and the Federal Ministry of the Republic of
Austria for Climate Action, Environment, Mobility, Innovation and
Technology (BMK) via the Austrian Space Applications Programme
with grant numbers 899537, 900565, and 911971.

Funding for the DES Projects has been provided by the U.S. Department of Energy, the U.S. National Science Foundation, the Ministry of Science and Education of Spain, 
the Science and Technology Facilities Council of the United Kingdom, the Higher Education Funding Council for England, the National Center for Supercomputing 
Applications at the University of Illinois at Urbana-Champaign, the Kavli Institute of Cosmological Physics at the University of Chicago, 
the Center for Cosmology and Astro-Particle Physics at the Ohio State University,
the Mitchell Institute for Fundamental Physics and Astronomy at Texas A\&M University, Financiadora de Estudos e Projetos, 
Funda{\c c}{\~a}o Carlos Chagas Filho de Amparo {\`a} Pesquisa do Estado do Rio de Janeiro, Conselho Nacional de Desenvolvimento Cient{\'i}fico e Tecnol{\'o}gico and 
the Minist{\'e}rio da Ci{\^e}ncia, Tecnologia e Inova{\c c}{\~a}o, the Deutsche Forschungsgemeinschaft and the Collaborating Institutions in the Dark Energy Survey. 

The Collaborating Institutions are Argonne National Laboratory, the University of California at Santa Cruz, the University of Cambridge, Centro de Investigaciones Energ{\'e}ticas, 
Medioambientales y Tecnol{\'o}gicas-Madrid, the University of Chicago, University College London, the DES-Brazil Consortium, the University of Edinburgh, 
the Eidgen{\"o}ssische Technische Hochschule (ETH) Z{\"u}rich, 
Fermi National Accelerator Laboratory, the University of Illinois at Urbana-Champaign, the Institut de Ci{\`e}ncies de l'Espai (IEEC/CSIC), 
the Institut de F{\'i}sica d'Altes Energies, Lawrence Berkeley National Laboratory, the Ludwig-Maximilians Universit{\"a}t M{\"u}nchen and the associated Excellence Cluster Universe, 
the University of Michigan, NSF NOIRLab, the University of Nottingham, The Ohio State University, the University of Pennsylvania, the University of Portsmouth, 
SLAC National Accelerator Laboratory, Stanford University, the University of Sussex, Texas A\&M University, and the OzDES Membership Consortium.

Based in part on observations at NSF Cerro Tololo Inter-American Observatory at NSF NOIRLab (NOIRLab Prop. ID 2012B-0001; PI: J. Frieman), which is managed by the Association of Universities for Research in Astronomy (AURA) under a cooperative agreement with the National Science Foundation.

The DES data management system is supported by the National Science Foundation under Grant Numbers AST-1138766 and AST-1536171.
The DES participants from Spanish institutions are partially supported by MICINN under grants PID2021-123012, PID2021-128989 PID2022-141079, SEV-2016-0588, CEX2020-001058-M and CEX2020-001007-S, some of which include ERDF funds from the European Union. IFAE is partially funded by the CERCA program of the Generalitat de Catalunya.

We  acknowledge support from the Brazilian Instituto Nacional de Ci\^encia
e Tecnologia (INCT) do e-Universo (CNPq grant 465376/2014-2).

This manuscript has been authored by Fermi Research Alliance, LLC under Contract No. DE-AC02-07CH11359 with the U.S. Department of Energy, Office of Science, Office of High Energy Physics.

The South Pole Telescope program is supported by the National Science Foundation (NSF) through the Grant No. OPP-1852617 and 2332483. Partial support is also provided by the Kavli Institute of Cosmological Physics at the University of Chicago.
Work at Argonne National Lab is supported by UChicago Argonne LLC, Operator of Argonne National Laboratory (Argonne). Argonne, a U.S. Department of Energy Office of Science Laboratory, is operated under contract no. DE-AC02-06CH11357.

This work is based on observations made with the NASA/ESA {\it Hubble Space Telescope}, using imaging data from the SPT follow-up GO programs 12246 (PI: C.~Stubbs), 12477 (PI: F.~W.~High), 13412 (PI: T.~Schrabback), 14252 (PI: V.~Strazzullo), 14352 (PI: J.~Hlavacek-Larrondo), and 14677 (PI: T.~Schrabback).
STScI is operated by the Association of Universities for Research in Astronomy, Inc. under NASA contract NAS 5-26555.
It is also based on observations made with ESO Telescopes at the La Silla Paranal Observatory under programs 086.A-0741 (PI: Bazin), 088.A-0796 (PI: Bazin), 088.A-0889 (PI: Mohr), 089.A-0824 (PI: Mohr), 0100.A-0204 (PI: Schrabback), 0100.A-0217 (PI: Hern\'andez-Mart\'in), 0101.A-0694 (PI: Zohren), and 0102.A-0189 (PI: Zohren).
It is also based on observations obtained at the Gemini Observatory, which is operated by the Association of Universities for Research in Astronomy, Inc., under a cooperative agreement with the NSF on behalf of the Gemini partnership: the National Science Foundation (United States), National Research Council (Canada), CONICYT (Chile), Ministerio de Ciencia, Tecnolog\'{i}a e Innovaci\'{o}n Productiva (Argentina), Minist\'{e}rio da Ci\^{e}ncia, Tecnologia e Inova\c{c}\~{a}o (Brazil), and Korea Astronomy and Space Science Institute (Republic of Korea), under programs 2014B-0338 and	2016B-0176 (PI: B.~Benson).
\appendix

%===========================================================
\section{Varying $a_{\rm dark}$ and $f_{\rm IDM}$}
\label{sec:a&f}
%===========================================================

The main results presented in this work are based on considering the tight coupling limit (meaning ${\rm log}_{10}[a_{\rm dark}/{\rm Mpc^{-1}}]>8$) and fixing the value of $f_{\rm IDM}=0.1$. This choice is motivated by the observation that the halo mass function (HMF) is mostly sensitive to the value of the DR temperature within the galaxy cluster mass range probed by SPT, as explained in~\cite{Mazoun:2023kid}. Nevertheless, for completeness, we also perform an analysis where $a_{\rm dark}$ and $f_{\rm IDM}$ are kept as free parameters. We include a flat prior $\mathcal{U}(0.0, 10.0)$ for ${\rm log}_{10}\,[a_{\rm dark}/{\rm Mpc}^{-1}]$, and $\mathcal{U}(0.001, 1.0)$ for $f_{\rm IDM}$.

\begin{figure}
    \centering
   \includegraphics[width=0.98\columnwidth]{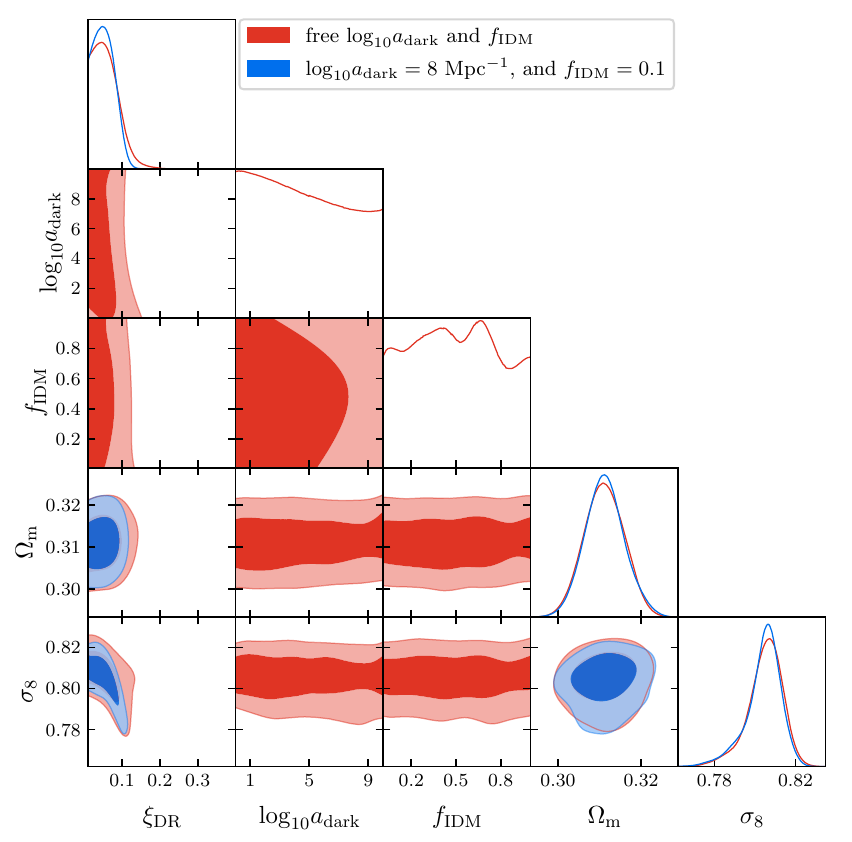}
    \caption{Posteriors of IDM$-$DR model parameters obtained from SPT-clusters$\times$WL+CMB+BAO data when allowing for variations in $a_{\rm dark}$ and $f_{\rm IDM}$ (red), compared to the case discussed in the main text where both parameters are fixed (blue).}
    \label{fig:a_and_f}
\end{figure}

As shown in Fig.~\ref{fig:a_and_f}, cluster abundance data as well as CMB+BAO data are only very weakly sensitive to $a_{\rm dark}$ and $f_{\rm IDM}$. However, importantly, we find that the constraint on the DR temperature $\xi_{\rm DR}$ is consistent with the results obtained when fixing the other two parameters as presented in Sec.~\ref{sec:IDM$-$DR}. This implies that the constraints on $\xi_{\rm DR}$ are robust with respect to variations in the ETHOS parameters $a_{\rm dark}$ and $f_{\rm IDM}$ over a wide range. Moreover, interestingly, the slight preference for non-zero $\xi_{\rm DR}$ observed in the main text is also present when allowing $a_{\rm dark}$ and $f_{\rm IDM}$ to vary. 

We note that we additionally performed an analysis for which we fixed ${\rm log}_{10}\,[a_{\rm dark}/{\rm Mpc}^{-1}]=8$, i.e. within the tight coupling limit, but included $f_{\rm IDM}$ as an additional free parameter, finding very similar results to those shown in Fig.~\ref{fig:a_and_f}. 

%===========================================================
\section{Full triangle plot}
\label{app:full_plot}
%===========================================================

\begin{figure*}
    \centering
    \includegraphics[width=0.99\linewidth]{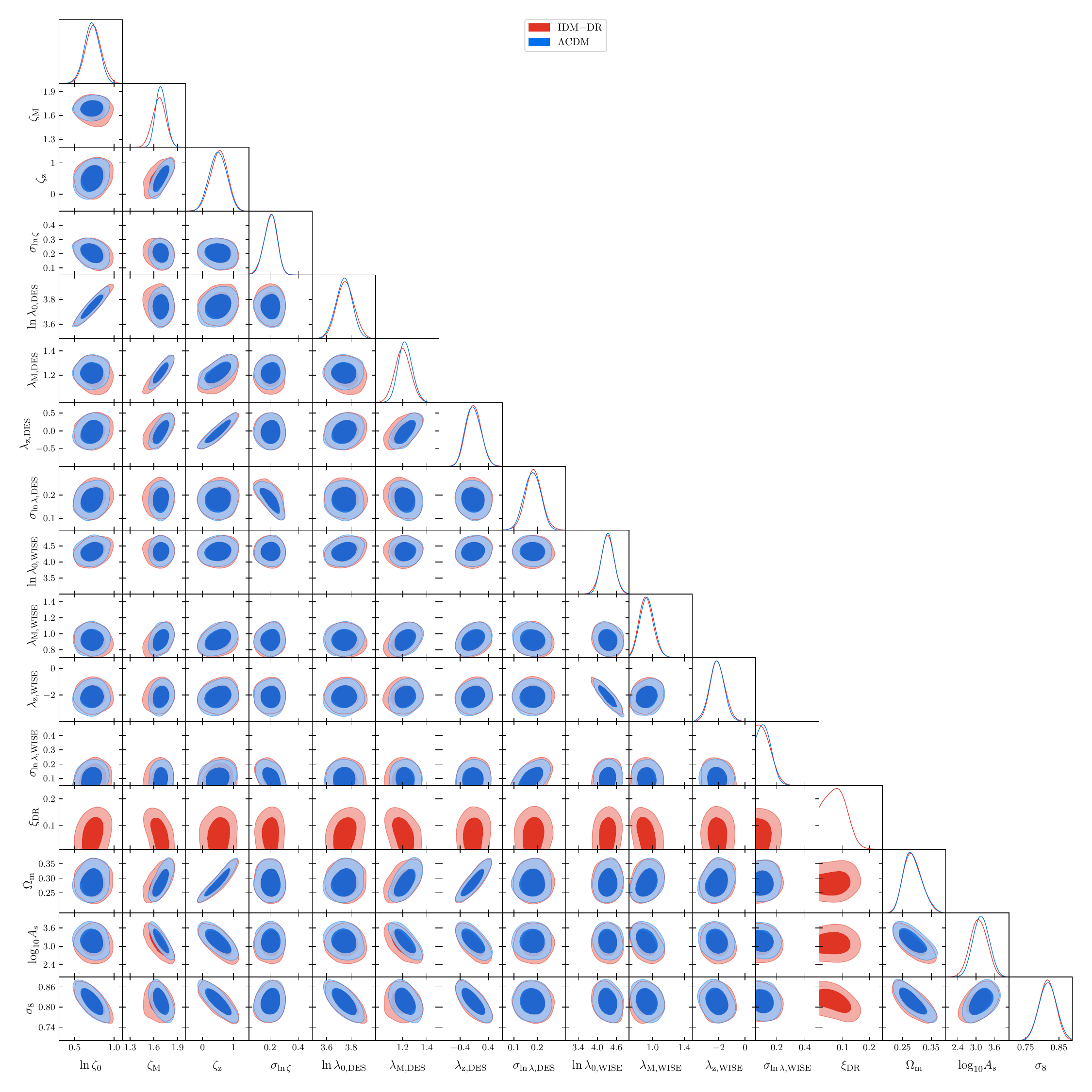}
    \caption{Posterior distribution of all parameters with a flat prior employed in the analysis of SPT-clusters$\times$WL within IDM$-$DR model (red) and \LCDM\ model (blue) from~\cite{Bocquet2024SPTcosmo}. }
    \label{fig:fullplot}
\end{figure*}

In Fig.~\ref{fig:fullplot} we show an extended version of Fig.~\ref{fig:results} for the constraints obtained from SPT-clusters$\times$WL data within the IDM$-$DR model. Fig.~\ref{fig:fullplot} includes all cosmological model parameters as well as the full set of parameters with flat-priors entering the cluster abundance and weak-lensing mass-calibration likelihoods that are jointly varied in our main analysis setup. The plot also shows consistent results with \LCDM\ (contours from~\cite{Bocquet2024SPTcosmo}).

%\bibliography{refs}

\end{document}